\documentclass[traditabstract]{aa} 
\usepackage{graphicx}
\usepackage{natbib}
\bibpunct{(}{)}{;}{a}{}{,} 
\usepackage{txfonts}
\def\OII{[\ion{O}{ii}]}
\def\OIII{[\ion{O}{iii}]}
\def\NeIII{[\ion{Ne}{iii}]}

\def\NII{[\ion{N}{ii}]}
\def\SII{[\ion{S}{ii}]}
\def\MgII{\ion{Mg}{ii}}

\begin{document}
   \title{GRB\,091127/SN\,2009nz and the VLT/X-shooter spectroscopy of its host galaxy: probing the faint end of the mass-metallicity relation\thanks{Based on observations made with ESO Telescopes at Paranal Observatory under programmes ID 084.A-0260 and 086.A-0874. }}


   \author{S. D. Vergani
          \inst{1,2,3}
          \and
          H. Flores
          \inst{2}
           \and
             S. Covino
             \inst{1}
              \and
             D. Fugazza
             \inst{1}
              \and
             J. Gorosabel
             \inst{4}
              \and
          A. J. Levan
            \inst{5}
            \and
          M. Puech
          \inst{2}
              \and
             R. Salvaterra
             \inst{6}
             \and
          J. C. Tello
          \inst{4}
              \and
             A. de Ugarte Postigo
             \inst{7}
             \and
             P. D'Avanzo
          \inst{1}
             \and
             V. D'Elia
          \inst{8,9}
           \and
          M. Fern\'andez
          \inst{4}
           \and
          J.P.U. Fynbo
          \inst{7}
           \and
             G. Ghirlanda
          \inst{1}
          \and
          M. Jel\'inek
          \inst{4}
          \and
          A.~Lundgren
          \inst{10,11}
          \and
          D. Malesani
          \inst{7}
           \and
          E. Palazzi
          \inst{12}
           \and
             S. Piranomonte
          \inst{8}
           \and
             M. Rodrigues         
              \inst{10}
          \and
          R. S\'anchez-Ram\'{\i}rez
          \inst{4}
          \and
          V. Terr\'on
          \inst{4}
          \and
             C.C. Th\"{o}ne
          \inst{4}
           \and
          L.A. Antonelli
          \inst{8}
          \and
             S. Campana
          \inst{1}
          \and
          A.J. Castro-Tirado
                    \inst{4}
          \and
          P. Goldoni
          \inst{3,13}
           \and
          F. Hammer
          \inst{2}
          \and
          J. Hjorth
          \inst{7}
          \and
          P. Jakobsson
          \inst{14}
          \and
          L. Kaper
          \inst{15}
          \and
             A. Melandri
          \inst{1}
       \and
          B. Milvang-Jensen
          \inst{7}
          \and
          J. Sollerman
          \inst{16}
          \and
             G. Tagliaferri
          \inst{1}
          \and
          N. R. Tanvir
          \inst{17}
          \and
          K. Wiersema
          \inst{17}
          \and
          R.A.M.J. Wijers 
          \inst{15}
          }

   \institute{INAF, Osservatorio Astronomico di Brera,
              via E. Bianchi 46, 23807 Merate, Italy\\
              \email{susanna.vergani@brera.inaf.it}
         \and
             GEPI-Observatoire de Paris Meudon. 5 Place Jules Jannsen, F-92195, Meudon, France
             \and
                    Laboratoire Astroparticule et Cosmologie, 10 rue A. Domon et L. Duquet, 75205 Paris Cedex 13, France
                    \and
                    Instituto de Astrof\'{\i}sica de Andaluc\'{\i}a (IAA-CSIC), Glorieta de la Astronom\'ia s/n, 18008 Granada, Spain
                    \and
                    Department of Physics, University of Warwick, Coventry, CV4 7AL, UK
                    \and
                    Dipartimento di Fisica e Matematica, Universit\`a dell'Insubria, via Valleggio 7, 22100 Como, Italy
                    \and
                    Dark Cosmology Centre, Niels Bohr Institute, University of Copenhagen, Juliane Maries Vej 30, 2100 Copenhagen \O, Denmark
                                        \and
                  INAF, Osservatorio Astronomico di Roma, via di Frascati 33, 00040 Monte Porzio Catone, Rome, Italy
                    \and
                    ASI-Science Data Center, via Galileo Galilei, 00044 Frascati, Italy
                    \and
                    European Southern Observatory, Alonso de C\'ordova 3107, Vitacura, Casilla 19001, Santiago 19, Chile
                    \and
                    Joint ALMA Observatory, Alonso de C\'ordova 3107, Vitacura - Santiago, Chile
                    \and
                    INAF, IASF di Bologna, via Gobetti 101, 40129 Bologna, Italy
                                        \and
                    Service d'Astrophysique, DSM/IRFU/SAp, CEA-Saclay, 91191 Gif-sur-Yvette, France
                     \and
                    Centre for Astrophysics and Cosmology, Science Institute, University of Iceland, Dunhagi 5, 107 Reykjav\'ik, Iceland
                     \and
                    Astronomical Institute Anton Pannekoek, University of Amsterdam,
    Science Park 904, 1098 XH Amsterdam, The Netherlands
                                       \and
                    The Oskar Klein Centre, Department of Astronomy, AlbaNova, Stockholm University, 106 91 Stockholm, Sweden
                    \and
                    Department of Physics and Astronomy, University of Leicester, University Road, Leicester LE1 7RH, UK
             }

   \date{Received 19 July 2011 / Accepted 16 September 2011}

 
\abstract{We perform a detailed study of the gamma-ray burst GRB\,091127/SN\,2009nz host galaxy at $z=0.490$ using the VLT/X-shooter spectrograph in slit and integral-field unit (IFU). From the analysis of the optical and X-ray afterglow data obtained from ground-based telescopes and {\it Swift}-XRT we confirm the presence of a bump associated with SN\,2009nz and find evidence of a possible jet break in the afterglow lightcurve. The X-shooter afterglow spectra reveal several emission lines from the underlying host, from which we derive its integrated properties. These are in agreement with those of previously studied GRB-SN hosts and, more generally, with those of the long GRB host population. We use the Hubble Space Telescope and ground based images of the host to determine its stellar mass ($M_{\star}$). Our results extend to lower $M_{\star}$ values the {\it M-Z} plot derived for the sample of long GRB hosts at $0.3<z<1.0$ adding new information to probe the faint end of the {\it M-Z} relation and the shift of the LGRB host {\it M-Z} relation from that found from emission line galaxy surveys. 
Thanks to the IFU spectroscopy we can build the 2D velocity, velocity dispersion and star formation rate (SFR) maps. They show that the host galaxy has a perturbed rotation kinematics with evidence of a SFR enhancement consistent with the afterglow position.

} 

\titlerunning{GRB\,091127 and its host galaxy}

   \maketitle
%

\section{Introduction}

The association of long gamma-ray bursts (LGRBs) with broad-lined type Ic supernovae (SNe) is now well established (e.g. \citealt{Hjorth2011} and references therein; see however GRB\,060614: \citealt{Della-Valle2006,Fynbo2006,Gal-Yam2006}). SN spectral features have been found in the afterglow spectra of 7 LGRBs (see \citealt{Starling2011} and references therein; \citealt{Sparre2011} and the more recent claim by \citealt{Berger2011} for SN\,2009nz). The association with a SN can also be inferred from the presence of a rebrightening of the GRB afterglow optical lightcurve after a few days from the burst. The first evidences of such bumps were reported by \cite{Bloom1999}, \cite{Galama2000} and \cite{Lazzati2001a}, and the spectroscopic
observations obtained for some of them confirmed the association (e.g. \citealt{Della-Valle2003}). 
The possibility that a massive collapsing star might launch relativistic jets capable of powering a gamma-ray burst was already predicted theoretically by \cite{Woosley1993} with the now called \textit{collapsar model} \citep{MacFadyen1999}. Further evidence for the LGRB -  SN Ic connection comes from the study by \cite{Kelly2008} showing that SNe Ic and LGRBs are similarly distributed within their hosts. 

SNe Ic, in particular the highly energetic, broad-lined Ic SNe, are 
likely produced by massive, rotating Wolf-Rayet (WR) progenitors (see \citealt{Crowther2007} for a review), making this kind of stars the best candidate LGRB progenitors (as single stars or in binary systems). Indeed WR spectral features have been found in the spectra of some LGRB host galaxies \citep{Hammer2006,Han2010}, nonetheless, the spatially resolved observations performed for one of these hosts (GRB\,980425) showed that the WR region was located hundreds of parsec away from the LGRB site \citep{Hammer2006}. WR wind signatures might also be found in the afterglow spectra \citep{van-Marle2008} even if up to date their claimed detections are debated \citep{Schaefer2003,Mirabal2003,Fiore2005,Starling2005a,Castro-Tirado2010,Chen2007}. 

The ratio of the LGRBs and SN Ib/c rate is of about 1\% \citep{Guetta2007,Soderberg2010}, therefore only a very small fraction of massive stars dies producing a 
LGRB. It is still not clear which are the peculiar conditions that lead a massive 
star to have the special kind of core collapse that triggers the formation of a jet and a LGRBs. Metallicity is one of the fundamental parameters predicted to impact the evolution of massive stars as well as their explosive deaths and, together with rapid rotation, is expected to play a fundamental role in the formation of LGRBs (e.g. \citealt{Heger2003,Yoon2010} and references therein). \cite{Modjaz2008} show that the SNe associated with LGRB seem to prefer lower metallicity environments than broad-lined Ic SNe without a LGRB association.

A possible way to retrieve information on LGRB progenitors and on the physical properties of LGRB regions is through spectroscopy of their host galaxies. Global properties of the hosts (metallicity, star formation rate, etc.) can be determined and more detailed investigations can be performed for nearby galaxies for which, through multi-slit and/or integral field unit (IFU) spectroscopy, we can build metallicity, SFR, density and velocity maps. It is therefore possible to look for peculiarities of the LGRB and its near-by regions and to study the kinematics of the gas in the galaxy. To date, spatially resolved studies of LGRB host galaxies have been performed only for the low-redshift GRB\,980425, GRB\,060505 and GRB100316D ($z=0.0085, 0.0889, 0.1218, 0.0593$, respectively; \citealt{Hammer2006,Christensen2008,Thone2008a,Levesque2011}). 

The study of LGRB hosts can also bring useful information to galaxy evolution studies. LGRB hosts form a sample of galaxies not selected by luminosity (the possibility of studying their properties is of course observationally partially biased towards brighter hosts though), that can be complementary to those of current surveys of galaxies. It has been recently shown that they are systematically offset to lower metallicities with respect to the mass-metallicity relation found from the surveys of emission line galaxies \citep{Han2010,Levesque2010,Mannucci2011}. In order to explain this behaviour and to build a complete picture of galaxy evolution, it is important to increase the LGRB host sample to confirm this result and to determine if and how it evolves at higher redshift. IFU surveys of intermediate redshift galaxies performed in the past (e.g.\,\citealt{Flores2006,Puech2006}) show the power of this technique in determining the dynamical properties of the galaxies demonstrating the presence of several galaxies with perturbed kinematics due to mergers or outflows and therefore adding important pieces to the galaxy evolution scenario. With IFU observations of LGRB host galaxies it will be possible to extend these studies to lower mass and lower metallicity objects.

Together with the analysis of the afterglow lightcurve, we present in this paper a detailed study of the host galaxy of GRB\,091127 at $z=0.49044$ associated with SN\,2009nz \citep{Cobb2010,Berger2011}, carried out using VLT/X-shooter \citep{DOdorico2004} echelle slit and IFU spectroscopy. This is the first paper presenting data obtained in the X-shooter IFU mode. 

In Section 2 we present the multiwavelength photometry of the GRB\,091127 afterglow and host galaxy obtained with ground-based facilities and the Hubble Space Telescope (HST). 
 These data and the {\it Swift}-XRT ones are fitted together to analyze the lightcurve (Section 3). 
 Sections 4 and 5 are dedicated to the longslit and IFU spectroscopic data, respectively. The results on the integrated properties of the host galaxy are reported in Section 6, whereas in Section 7 the IFU 2D host galaxy maps are shown. Our findings are discussed in Section 8 and conclusions are drawn in Section 9.

\section{Photometry of the afterglow and of the host galaxy}

\label{photometry}

We obtained multiwavelength photometry of the afterglow lightcurve and the burst host galaxy from $\sim1$ hr to $\sim400$ days (observer frame) after the GRB explosion, using the HST and several ground based facilities (see Tab.\,\ref{phot}; Fig.\, \ref{lc}).

\begin{table}[]
\small
\caption{GRB\,091127 OT and host galaxy photometry }
\begin{tabular}{lllll}
\hline\hline
Telescope &Obs. time & Filter & Mag$^{(a)}$ & Exp. \\
& (days) & && time (s)\\
\hline
TNG		&	368.946	&	{\it B}	&	$24.25\pm0.18$	&	1200\\
GTC		&	8.94568	&	{\it g}	&	$23.03\pm0.10$	&	90\\
GTC		&	34.9434	&	{\it g }  &	$23.38\pm0.38$	&	30\\
GTC		&	256.219	&	{\it g}   &	$24.23\pm0.26$    &	840\\
NOT		&	48.8811 &	{\it V  }	&	$23.43\pm0.05$	&	3200\\
TNG		&	368.963	&	{\it V}	&	$23.73\pm0.18$	&	1200\\
TNG &5.95030	&	{\it R}	&	$21.75\pm0.11$&960\\
OSN&6.95917	&	{\it R	}&	$21.94\pm0.18$&	1800\\
GTC		&	8.94099	&	{\it r}	&	$22.21\pm0.06$	&	90\\
GTC		&	34.9392 &{\it 	r}	&	$22.52\pm0.11$	&	30\\
GTC		&	256.230 &{\it 	r }  &	$23.15\pm0.13$    &	600\\
1.23mCAHA &	0.03599&	{\it I} &	$15.28\pm0.35$  &	120\\
1.23mCAHA &	0.06309 & 	{\it I }& $16.40\pm0.05$  &   120\\
NOT	&	48.8440 &{\it 	I }& $22.51\pm0.08$ &   720\\
TNG	&	368.978 &{\it 	I} &	$22.77\pm0.19$ &	900\\
GTC		&	8.95059	&	{\it i	}&	$22.00\pm0.07$	&	90\\
GTC		&	34.9407	&	{\it i	}&	$22.03\pm0.06$	&	30\\
GTC		&	256.241	&	{\it i  } &$	22.86\pm0.11 $   &	840\\
GTC		&	8.95529 &{\it 	z}	&	$22.17\pm0.11$	&	90\\
GTC   	&	34.9767 &{\it 	z}   & 	$21.97\pm0.08	$&	30\\
GTC 	         &	256.254 &{\it 	z } 	&	$22.39\pm0.28$    &	810\\
REM	&	0.05176 &{\it 	J}	&	$15.23\pm0.18$	&	150\\
REM	&	0.06481	&	{\it J	}&	$15.31\pm0.14$	&	300\\
REM	&	0.04730	&	{\it H	}&	$15.15\pm0.18$	&	150\\
REM	&	0.05692	&	{\it H	}&	$14.90\pm0.11$	&	300\\
HST &  383.154&{\it H} & $21.62\pm0.04$&906\\
3.5mCAHA &	4.93481	&	{\it K}	&	$19.45\pm0.17$	&	1895.4\\
\hline
\hline

\end{tabular}

(a) Magnitudes are corrected for Galactic extinction of $E(B -V)=0.038$\,mag \citep{Schlegel1998}.
{\it B}, {\it V}, {\it R}, {\it I}, {\it J}, {\it H}, {\it K} are Vega magnitudes, whereas {\it g}, {\it r}, {\it i} and {\it z} are AB magnitudes. 
\label{phot}
\end{table}

The early epoch $J$ and $H$ images, observed with REM-REMIR\footnote{http://www.rem.inaf.it/}, were automatically reduced by the task PREPROCESS (\citealt{di-Paola2001}; sky subtraction, average of 5 dithered images). The photometry was performed with GAIA\footnote{http://astro.dur.ac.uk/$\sim$pdraper/gaia/gaia.html} using an aperture of 5\arcsec\,(5 pixel) compatible with the measured FWHM (2.5\arcsec). The zero point calibration is based on three 2MASS stars \citep{Skrutskie2006} sufficiently bright in the GRB field.
The TNG-DOLoRes\footnote{http://www.tng.iac.es/} images, 16 in the $R$ filter for the afterglow observation, 2 in the filters $B$, $V$ and I for the late time host observation, were shifted and averaged. The fringes in the $I$ images were removed using the frame of  reference provided by the TNG web pages. The photometry was performed with GAIA using the PSF method (the measured seeing is $\sim1\arcsec$ in all cases). The calibration of the zero point is based on 9 secondary standard stars in the GRB field. 

The reduction of both the optical and near-IR GTC\footnote{http://www.gtc.iac.es/en/pages/gtc.php}, NOT\footnote{http://www.not.iac.es/} and CAHA\footnote{http://www.caha.es/} data was done with IRAF\footnote{IRAF is a software distributed by \textit{National Optical Astronomy Observatories}.} following standard procedures. 
The photometry of the reduced images was carried out running PHOT aperture photometry as implemented in IRAF.  
We used as radius of the aperture the FWHM of the seeing.
The  optical calibration was obtained observing a standard star at an airmass similar to the one of the GRB and assuming the extinction coefficients of La Palma. The $K$-band calibration is based on 12 bright unsaturated field stars present in the 2MASS catalogue.


Continuum observations at 870 $\mu$m were also carried out, using the LABOCA bolometer array, installed on the Atacama Pathfinder EXperiment (APEX\footnote{APEX is a collaboration between the Max-Plank-Institut f\"ur Radioastronomie, the European Southern Observatory and the Onsala Space Observatory}) telescope, on 2009 November 28th and 29th. The total on source integration time of the two combined epochs was 6.4\,hr (2.9 and 3.5\,hr in the first and second epoch, respectively). Data were reduced using the Crush2 software\footnote{http://www.submm.caltech.edu/$\sim$sharc/crush/}. We obtained 3$\sigma$ upper limits of 14.9 mJy and 13.2mJy for each of the two epochs, respectively, and 9.6 mJy for the combined epochs.

Host galaxy observations of GRB 091127 using the
HST have been obtained on 16 December 2010, utilizing
the WFC3 in the $H$ band (F160W). An exposure time of 906\,s was obtained
in two dithered exposures, and reduced in the standard HST fashion.
At the location of the optical afterglow we clearly see the underlying
host galaxy, whose centroid is offset relative to the position of
the optical afterglow by only $\sim$ 0.2\arcsec, confirming the analysis of \cite{Cobb2010}. The host has
$H$(AB) = $22.88 \pm 0.04$, $H$(Vega) = $21.62 \pm 0.04$, that, considering the optical host galaxy magnitudes obtained with the TNG, suggests
that, like many LGRB hosts the host of GRB~091127 is relatively blue. 

We calculated the probability $P_{\rm ch}$ that the identified galaxy is a chance superposition and not physically related to the GRB. Using the procedure described in \cite{Bloom2002}, we find $P_{\rm ch}\sim 4\times10^{-3}$.

\begin{table}[!h]
\begin{center}
\small
\caption{Fit results. Error ranges for the optical analysis and the X-ray temporal analysis are at 1$\sigma$, whereas for the X-ray spectral analysis they represent the 90\% confidence level.}
\label{fit}
\begin{tabular}{l|cc}
\hline\hline
\multicolumn{3}{c} {Optical data}\\
\hline
&Best fit&Range\\
\hline
$\alpha_1$ &0.51&[0.30, 0.73]\\
$\alpha_2$ & 1.74&[1.46, 2.08]\\
$t_{\rm break}$ (days)&0.45&	[0.24, 0.71]\\
$\beta$& +0.18&	[-0.43, 0.81]\\
$E(B-V)_{\rm LMC}$ &0.16	&[0.03, 0.31]\\
$k$&1.1	&[0.71, 1.48]\\
\hline
 $\chi^2$/d.o.f.  &\multicolumn{2}{c} {53.96/44=1.23}\\
 \hline
 \hline
 \multicolumn{3}{c} {X-ray data: temporal analysis}\\
 \hline
&Best fit&Range\\
\hline
$\alpha_1$ &1.09&[1.03, 1.14]\\
$\alpha_2$ & 1.55&[1.51, 1.58]\\
$t_{\rm break} $ (days)&0.38&	[0.29, 0.51]\\
\hline
 $\chi^2$/d.o.f.  &\multicolumn{2}{c} {376.09/364=1.03}\\
\hline
\hline
 \multicolumn{3}{c} {X-ray data: spectral analysis}\\
 \hline
&Best fit&Range\\
\hline
$N_{\rm H}$ ($10^{21}$ cm$^{-2}$) &1.31&[0.83,1.81]\\
$\beta$ & 0.93&[0.82,1.04]\\
\hline
 C-stat(d.o.f.)  &\multicolumn{2}{c} {327.42(395)}\\
\hline
\hline

\end{tabular}

\end{center}
\end{table}

\section{Analysis of the afterglow emission }

\begin{figure*}
 \centering
   \hspace{-1.0cm}
   \includegraphics[width=9cm, angle=270]{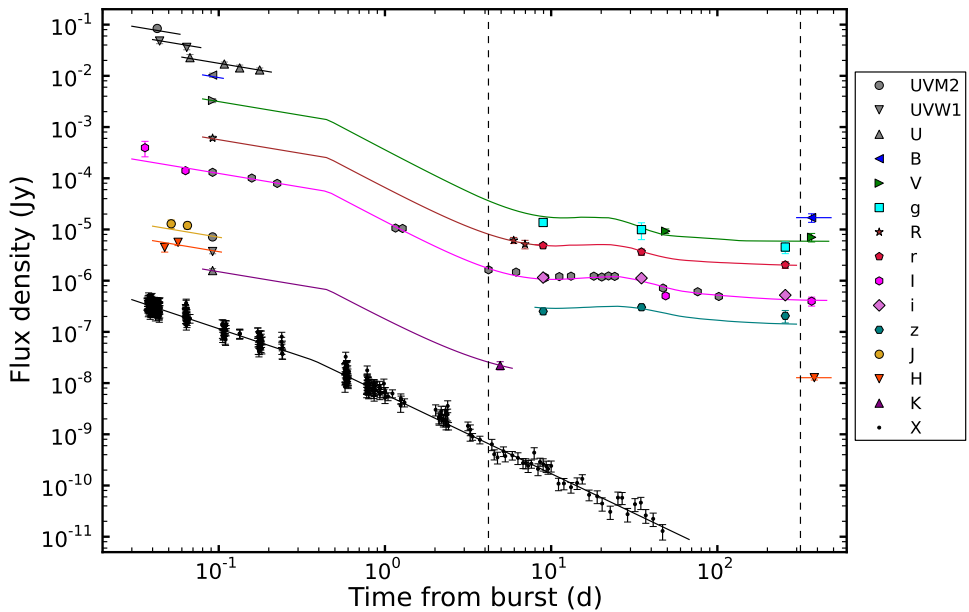}
      \caption{Optical and X-ray light-curves of the GRB\,091127 afterglow. The symbol legend is to the right of the plot. Fluxes have been rescaled for clarity, except for the $R$- and $r$-filters ($UVM2\times200$, $UVW1\times100$, $U\times50$, $B\times20$, $V\times6$, $g\times6$, $I\times0.2$, $i\times0.2$, $z\times0.05$, $J\times0.01$, $H\times0.005$, $K\times0.002$ and $X\times0.01$). Colored symbols correspond to our data set, whereas grey symbols refers to the data published in the literature (see \citealt{Cobb2010} and references therein). The solid curves show the best fit for each band. A color version of this figure is available online.
              }
         \label{lc}
   \end{figure*}

We added to the optical/near-infrared (NIR) observation presented in Sect.\,\ref{photometry} the X-ray afterglow products (see Fig.\,\ref{lc}) found in the {\it Swift}-XRT light-curve and spectra repositories \citep{Evans2007,Evans2009}.

In order to analyse the afterglow emission, we fitted the photometric data in Table \ref{phot} together with the data published by \cite{Cobb2010} and
the X-ray data with power-laws both for the spectral and temporal regimes,
following usual conventions ($F_\nu \propto t^{-\alpha} \nu^{-\beta}$). For the optical spectral analysis we also
considered rest-frame extinction following the Milky Way (MW), Large and Small
Magellanic Clouds (LMC and SMC), and starburst (SB) recipes \citep{Pei1992,Calzetti1994}. For the X-ray data we considered neutral rest-frame absorption \citep{Morrison1983}. In both cases the effect of extinction due to dust and
absorption due to gas in the Galaxy has been previously removed. For the optical
data fit we used a broken power law with the addition of a late-time constant component, the host galaxy, and a SN component modelled following SN\,1998bw \citep{Galama1998,McKenzie1999,Sollerman2000,Patat2001,Sollerman2002}, i.e. the same used by \cite{Cobb2010}. The SN template was K-corrected assuming a linear relation between
the adjacent bands of the SN\,1998bw observations. A possible scaling
factor $k$ for the whole SN component was also introduced in the fit.
The temporal properties of the SN turned out to be fully consistent in all bands with SN\,1998bw shifted at the redshift of GRB\,091127 and with $k=1.1$. It dominates the optical light-curve from about a week and it is significant up to
about 200 days from the burst, when the host galaxy stable emission began to be
observed. Recently \cite{Berger2011} claimed the detection of supernova features in their Gemini spectra, confirming the association of GRB\,091127 with a SN and finding that the SN shows spectral properties similar to those of XRF\,060218/SN\,2006aj. 

Although the global shape of the optical light-curve allowed us to separate well 
the various components, it immediately turned out that there are severe cross-calibration
uncertainties among the many instruments 
providing the available data. \cite{Cobb2010} mention a systematic calibration error of 0.05 considering their data alone. In order to obtain an
acceptable $\chi^2$ for the global fit we needed to add a 10\% uncertainty in
quadrature. This effectively means that we are not able to study possible low-amplitude
variations, and only the global light-curve shape can be effectively
modelled.

The X-ray light-curve is remarkably smooth and could be well modelled
by a broken power-law with a break at about 33\,ks. The break is consistent with
being simultaneous at optical wavelengths. The post-break decay slopes are
consistent between optical and X-rays within errors, while the pre-break decay in the
optical is significantly shallower than at higher energies. The X-ray spectral slope
is slightly lower than 1 while, within the weak constraints of the fit, the optical spectral slope is much harder and it shows no evidence of spectral evolution (see also
\citealt{Cobb2010}). This is not an unprecedented behaviour, as very blue optical
spectra have already been reported in the past (e.g. for GRB\,060908, \citealt{Covino2010}). The optical spectral slope and rest-frame extinction show a high degree of
degeneracy. The best fit would indeed require a non negligible rest-frame
absorption following the LMC extinction curve (see Tab. 2). All other recipes provided worse fits. \cite{Cobb2010} for their observation covering a shorter wavelength range find that a smaller amount of absorption ($A_V = 0.2$\,mag, corresponding to $E(B-V)_{\rm LMC} = 0.06$\,mag using their parameterization) is compatible with their data.

The results of the fits are shown in Tab.\,\ref{fit}.

\section{X-shooter long slit spectroscopy}

\begin{table*}[!htp]
\begin{center}
\small
\caption{
X-shooter observation log. From column 2 to 8 we report the observation date together with the observing run, the observation mode, the arms used, the corresponding binning, total exposure time, slit width and central wavelength resolution, for both slit and IFU observations.}
\label{log}

\begin{tabular}{cccccccc}
\hline\hline
GRB\,091127&   Date   & Mode &  Arm &Binning& Total exp. time & Slit width & Resolution \\
                 &                 Program ID   &   &         & & on source (s)          &   (\arcsec)     &      $\lambda/\Delta\lambda$        \\
\hline\hline
slit &  02 Dec. 2009   & Nodding  & UVB &$1\times2$&6000  &  1.0       & 5100         \\
								&084A-0260(C) &	& VIS  &$1\times2$&6000  &  0.9       & 8800         \\
           						                  && & NIR &$1\times1$ & 6000  &  0.9       & 5100          \\
						                  \hline
IFU & 10 Oct. 2010   & Staring& UVB&$1\times1$ & 3300  &  0.6       & 7900 \\  
								&086.A-0874(A)  & & VIS &$1\times1$& 3300  &  0.6       & 12600  \\
           							& & & NIR&$1\times1$&  3300  &  0.6       & 8100   \\
							\hline

\hline
\end{tabular}
\end{center}
\end{table*}

X-shooter long slit spectra of GRB\,091127 were obtained under the GTO program (084A-0260; PI: J. Fynbo) dedicated to GRB afterglow spectroscopy. For technical reasons, we could start the observation only about 4.2 days after the GRB explosion \citep{Thone2009}, at 03:17 UT on Dec 2, 2009. The observation consisted of $4\times1500$\,s exposures in the UVB, VIS and NIR arms (see Tab. \ref{log}) with an average seeing of $1\arcsec$. 

We processed the frames using version 1.2.0 of the X-shooter data
reduction pipeline \citep{Goldoni2006}. The data reduction was performed with the following steps. The raw frames were first subtracted and cosmic ray hits were detected and removed using the method developed by \cite{van-Dokkum2001}. 
The frames were then divided by a master flat field produced using daytime
flat field exposures with halogen lamps.
The spectral orders were extracted and rectified in wavelength space using a wavelength solution
previously obtained from calibration frames. The resulting rectified orders were then shifted and added to superpose them thus obtaining the sky subtracted 2D spectrum. After these steps, the orders were merged and in the overlapping regions the merging was weighted by the errors which were propagated during the process. From the resulting 2D merged spectrum, a one dimensional spectrum was extracted at
the source position with the corresponding error files
and bad pixel maps.

The afterglow continuum is detected across the whole spectral range, even if its S/N is pretty low (also due to the effect of the nearby Moon) reaching a maximum of 5.5 in the VIS arm of the spectrum.
Several emission lines from the host galaxy of GRB\,091127 are clearly identified, superimposed on the afterglow continuum, in the UVB and VIS spectra
at a common redshift of $0.49044\pm0.00008$ (see also \citealt{Thone2009,Cucchiara2009a}) and spanning about 150\,km s$^{-1}$. We do not detect any line in the NIR spectra. 
The overall 1D spectrum was flux calibrated using the reference standard stars observed during the same night and it was subsequently cross-checked and corrected using the corresponding magnitudes of the afterglow according to the photometric data and fit reported in Section \ref{photometry}. Slit-aperture flux loss corrections are therefore included in this procedure. 

The observed emission line fluxes were measured using a self developed code and cross-checked with the SPLOT task in IRAF. The results, corrected for Galactic extinction, are listed in Table\,\ref{lineflux}. 

We looked for absorption lines but we can only report a tentative \MgII\,$\lambda$2796 absorption at the limit of the detection, with a rest frame equivalent width of about 2.5\,\AA\, blueshifted by about 460\,km\,s$^{-1}$ from the emission lines. 

 \begin{figure}
   \includegraphics[width=8cm]{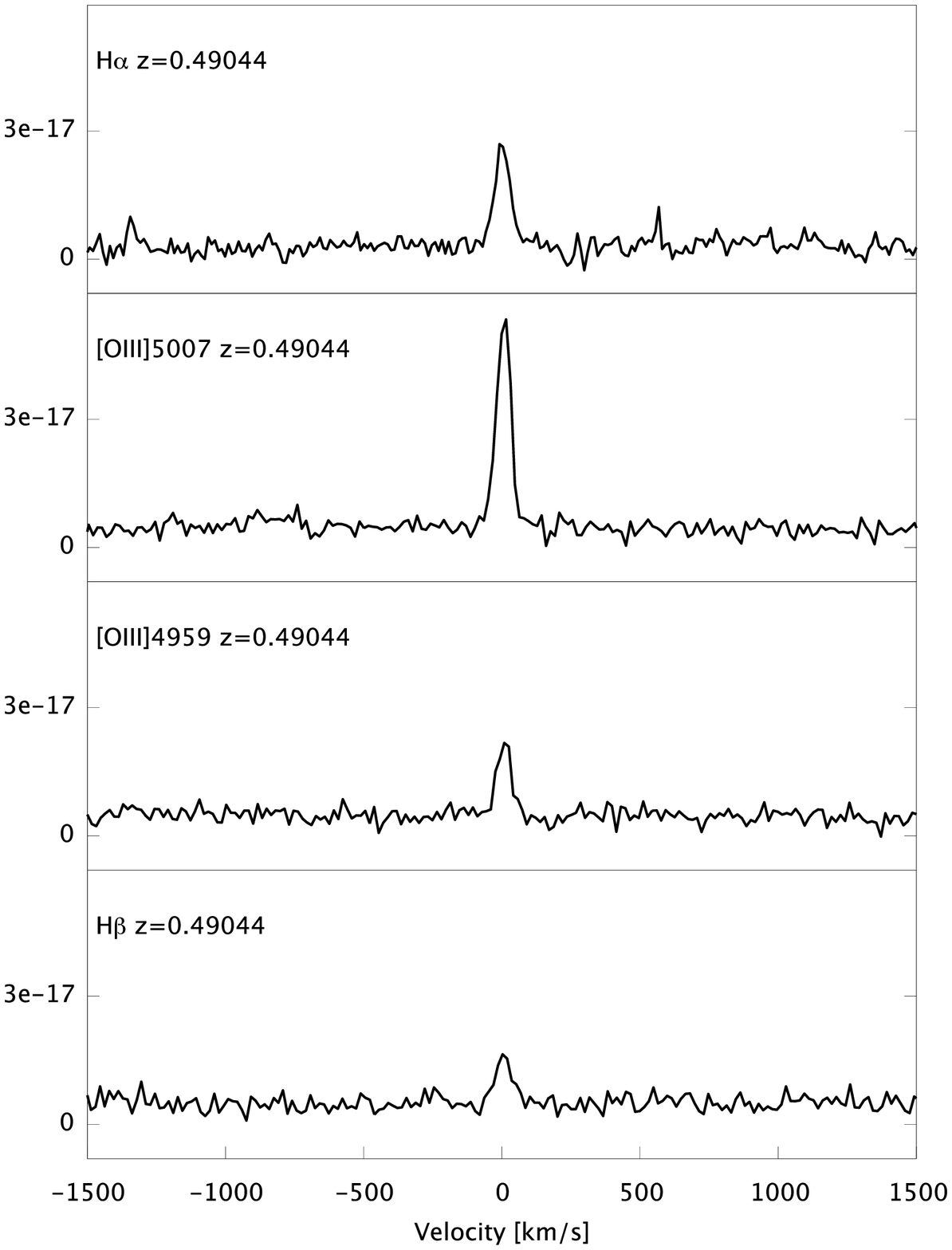}
     \includegraphics[width=8cm]{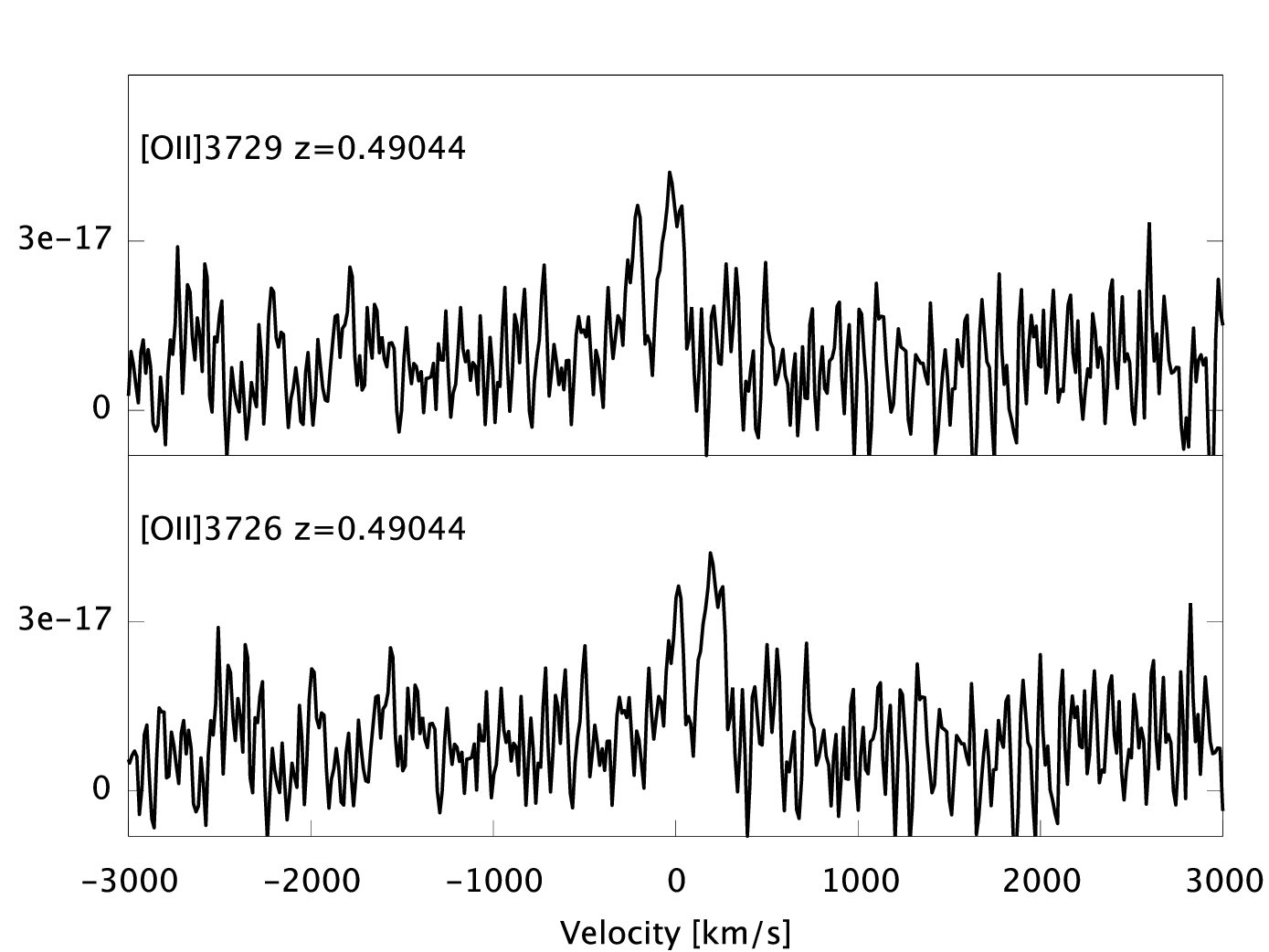}
      \caption{Emission lines in the 1D slit VIS and UVB spectra (upper and lower panel, respectively), shown in the velocity space centered at $z=0.49044$. Flux unit are ergs s$^{-1}$ cm$^{-2}$ $\AA^{-1}$.
              }
         \label{emission}
   \end{figure}

\begin{table}[!h]
\begin{center}
\small
\caption{Observed emission-line fluxes corrected for Galactic extinction. Upper limits are calculated at 3-$\sigma$. Errors shown here do not take into account a 10\% systematics due to flux calibration uncertainties.}
\begin{tabular}{lc}
\hline\hline
Emission line&  Flux\\
                 &       $10^{-17}$\,erg s$^{-1}$ cm$^{-2}$ \\ 
\hline\hline
\OII$\lambda$3726  &$5.0\pm1.2^{a}$\\
\OII$\lambda$3729 &$8.2\pm1.2^{a}$\\
\NeIII$\lambda$3869 &$<2.8$\\
H$\delta$   &$<2.5$\\
H$\gamma$   &$1.0\pm0.4$\\
\OIII$\lambda$4363 &$<1.6$\\
H$\beta$	&$2.0\pm0.3$\\  
\OIII$\lambda$4959 &$2.8\pm0.3$\\
\OIII$\lambda$5007 &$7.9\pm0.3$\\
H$\alpha$  &$5.5\pm0.3$\\
\NII$\lambda$6584  &$< 1.0$\\
\SII$\lambda$6717  &$1.5\pm0.5^{b}$\\
\SII$\lambda$6731  &$< 1.2$\\

\hline
\hline
\end{tabular}

\end{center}
 
\textit{Notes}: (a) The larger errors on the \OII\,fluxes are due to the nearby Moon and to the fact that the \OII\,lines fall right
in the range of the dichroic between the UVB and VIS arm where the transmission is therefore not optimal.
(b) Tentative detection: possible emission near a sky line.
\label{lineflux}

\end{table}

\section{X-shooter IFU spectroscopy}

\begin{figure}
   \centering
   \includegraphics[width=7cm, angle=270]{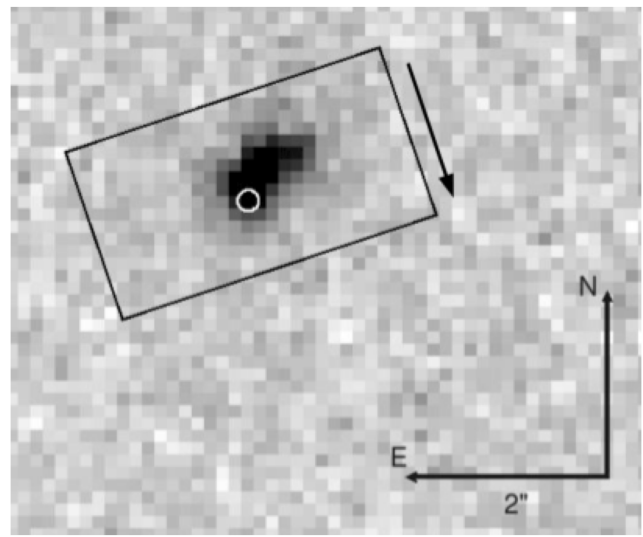}
      \caption{HST H-band (F160W) image of the field of GRB\,091127. The $1\farcs8\times3\farcs5$ IFU field of view is shown over the GRB host galaxy. The black arrow represents the direction of the IFU maps. The afterglow position (see Sect. 7.1) is indicated by the white circle. 
              }
         \label{tng}
   \end{figure}

X-shooter IFU spectra of GRB\,091127 were obtained under excellent seeing conditions in October 2010 within the Italian-French GTO program (086A-0874; PI: S. Piranomonte) dedicated to GRB host galaxy spectroscopy (see Tab. \ref{log}). We used an object-sky sequence with an exposure time of 3300\,s in each position. The position of the IFU field of view is shown in Fig.\,\ref{tng}. The IFU was positioned on the host galaxy centre using blind offsets. The pointing error should be less than 0.1\arcsec .

\subsection{IFU data reduction}

The IFU data reduction recipe is still under development within the ESO pipeline. We have been able however to reduce the VIS arm data, using the recipe developed for the slit observations and correcting for the sky background via standard programs under IRAF and IDL. Differently from the slit mode, for the IFU mode the atmospheric dispersion compensators (ADCs) are not in the optical path, therefore the IFU mode is strongly affected by the atmospheric dispersion. Specific IDL programs were used to correct the athmospheric dispersion and construct the datacube (see \citealt{Flores2011}). The standard star GD71 was used to calibrate the flux of the datacube.

The final datacube (constructed around the \OIII$\lambda$5007 emission line) has a spatial resolution of $0\arcsec .1\times0\arcsec .6$ (equivalent to $0.6\times 3.6$\,kpc$^2$) and we kept the spectral resolution of the observations ($R=12600$ for the VIS arm).

\section{Integrated properties of the host galaxy}

Using the emission lines present in the slit spectrum we can determine some integrated properties of the host galaxy. 
Assuming a temperature $T_e=10000$\,K, from the \OII$\lambda$$\lambda$3727,\,3729 and \SII$\lambda$6717,\,6731 ratio we infer an electron density $n_e\le200$\,cm$^{-3}$. The comparison of the observed flux ratio of the Balmer hydrogen emission lines with the values expected in the case B recombination in \cite{Osterbrock1989} shows that no significant Balmer absorption correction is needed and that the ratios are consistent with $A_V = 0$\,mag (within errors values of $A_V<0.3$\,mag are allowed), indicating a blue host galaxy dominated by a young star population as also supported by the galaxy colors . 

We determine the metallicity of the host using the $R_{23}$ calibrator by the \cite{Kobulnicky2004} method as described in \cite{Kewley2008}. 
We find 12+log(O/H)=$8.72\pm0.16$ using the lower branch formula and 12+log(O/H)=$8.04\pm0.16$ for the upper branch. This indicates that we are at the turnover point of the double-valued $R_{23}$ diagnostic, that in fact occurs at the maximum value of $\log(R_{23})=1$ (see Figs.\,6 and 7 of \citealt{Kobulnicky2004}) which corresponds to the value determined from our data. From the models, the metallicity corresponding to this value is 12+log(O/H)\,$\sim8.4$ \citep{Kobulnicky2004}, with an error of $\sim 0.1$\,dex.

Using the the relations found by \cite{Kewley2008} we can convert this value to the \cite{Pettini2004a} O3N2 diagnostic finding 12+log(O/H)$=8.2$. This result is consistent with the limit on the metallicity determined using the O3N2 diagnostic with our data, i.e. 12+log(O/H)$<8.3$ (it should be noted that this diagnostic is valid only for O3N2\,$<2$. We can only assume that this condition is satisfied for the GRB\,091127 host, since we have only an upper limit for the \NII\, flux giving O3N2\,$>1.2$). 

We also determined the metallicity using the method followed by \cite{Mannucci2010} based on the simultaneous fitting of strong line ratios described in \cite{Maiolino2008}. We find in this case 12+log(O/H)=$8.34^{+0.10}_{-0.20}$.

We calculate the SFR using the H$\alpha$ emission line following the prescription given in \cite{Kennicutt1998} and considering both the initial mass function (IMF) of \cite{Salpeter1955} and that proposed by \cite{Baldry2003} used by \cite{Savaglio2009} for GRB host galaxies. We find 
 SFR$_{\rm H\alpha}=0.40,0.22\pm0.03$\,M$_{\odot}$\,yr$^{-1}$, respectively.

We used the HST and TNG late time photometry of the GRB host to build the SED and infer the stellar mass ($M_\star$) of the host galaxy. 
We fit the SED using Charlot \& Bruzual models \citep{Bruzual2003,Bruzual2007}. We assumed a
Salpeter IMF and constructed grids of $\tau$-exponentially declining
star formation histories spaced at 0.1\,Gyr intervals from
$\tau=0$ to 0.5\,Gyr, and at 0.5\,Gyr intervals beyond,
and age $t$. Following the results obtained from the slit spectroscopy, we fixed $Z=0.008$ (that, among the values available, is the closest to the 12+log(O/H) metallicity determined above) and $A_V = 0$\,mag. 
Within a confidence interval of 68\% 
we find $\tau \in[100; 6000]$\,Myr, $t\in[227;905]$\,Myr and a stellar mass of  
$\log(M_\star/M_\odot)$=8.6$\pm0.1$. Considering the average conversion factor to derive stellar masses using the IMF adopted in \cite{Savaglio2009}, the mass we obtained corresponds to $\log(M_\star/M_\odot)$=8.3. \cite{Levesque2010} found that their masses determined using a \cite{Chabrier2003} IMF are consistent with those of \cite{Savaglio2009}, therefore this latter value can be used when comparing with the masses of the GRB host galaxies studied by \cite{Levesque2010}.

The specific star formation rate (SSFR) calculated with the SFR and $M_\star$ determined above is log(SSFR)(yr$^{-1}$)=-9.0$\pm0.1$.

From the host galaxy late time photometry we derive an absolute $B$-band magnitude of $M_B=-18.4$. The host galaxy is therefore sub-luminous even considering the luminosity function of irregular and starburst galaxies only \citep{Dahlen2005}.

\section{IFU 2D maps}

\subsection{Velocity and velocity dispersion maps}

We used the software developed for the FLAMES/GIRAFFE instrument to recover the velocity field and the velocity dispersion ($\sigma$) map \citep{Flores2006,Yang2008}. Velocity fields have been derived after subtraction of a $\sigma$-clipped mean of all velocities, and considering only spaxels with $\mathrm{S/N} > 4$. The fit of each accepted spaxel was checked by eye.

 \begin{figure}[!h]
   \centering
   \includegraphics[width=8cm, angle=270]{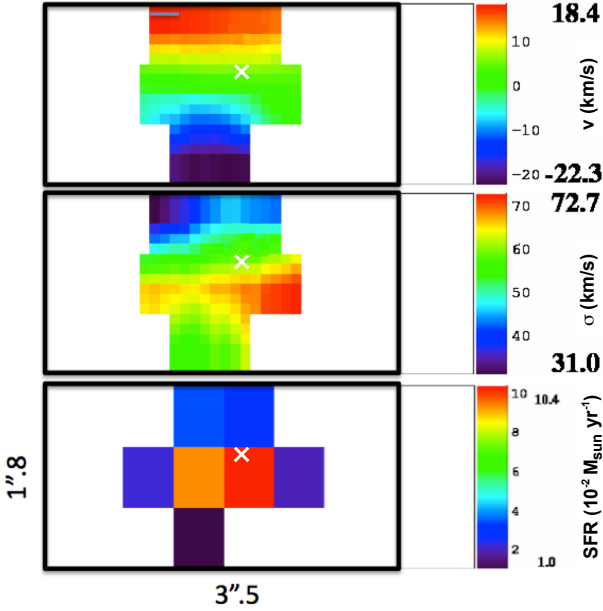}
      \caption{\textit{From top to bottom}: IFU velocity, velocity dispersion and SFR (Salpeter IMF) maps ($1\arcsec=6.018$\,kpc). Each map is orientated following the direction of the arrow reported in Fig.\,\ref{tng} (bottom to top). The afterglow position is indicated by the white cross. The velocity and velocity dispersion maps are interpolated to spaxels of $0\arcsec.1\times0\arcsec.1$ for better viewing. A color version of this figure is available online.}
         \label{vel}
   \end{figure}

Fig.\,\ref{vel}Ê shows the velocity field and the velocity dispersion ($\sigma$) maps obtained using the \OIII$\lambda5007$\AA\, emission line of the host galaxy. For better viewing we show the interpolated maps (spaxels of $0\farcs1\times0\farcs1$, real spaxels being $0\farcs1 \times0\farcs6$). The velocity map indicates a rotating galaxy and the $\sigma$-map shows an offset of the velocity dispersion peak from the galaxy center. Such kinematics have been defined as perturbed rotation by \cite{Flores2006} for their 3D VLT/GIRAFFE galaxy survey. Even using the HST image of the galaxy, it is difficult to check if the morphology and the kinematics have the same principal axis, preventing a deeper interpretation of both maps. We can only notice that the velocity field does not seem to follow the galaxy major axis direction. Galaxies with the velocity field not aligned with their major axis have been found in the 3D VLT/GIRAFFE survey and defined as galaxies with complex kinematics (e.g.: \citealt{Flores2006,Puech2006}).

Using the HST host galaxy image and the GEMINI afterglow images we could determine an offset for the afterglow position of $0.29\pm0.05$\arcsec south and $0.08\pm0.05$\arcsec east from the host galaxy centre (we find comparable results also using different afterglow images from our dataset). While the offset in the southern direction is consistent whit that found by \cite{Cobb2010}, they found a small offset in the western direction. The discrepancy could be due to the better resolution of the host galaxy of the HST image.
In both cases, the GRB afterglow is located at the border of the high dispersion region.

\subsection{SFR map}

 The SFR map has been reconstructed from the H$\alpha$ emission line assuming no extinction (as deduced from the slit observation). Given the weakness of theÊ H$\alpha $ emission line, we build a datacube with a spatial resolution of $0\farcs5 \times0\farcs6$ (equivalent to $3.0\times 3.6$\,kpc$^2$), in order to increase the S/N of each spaxel. The emission of each spaxel was measured with automatic software and manually with IRAF.  Only spaxels with a S/N $\ge$ 3 are considered in the map shown in Fig.\,\ref{vel}. The peak of the star formation is close to the position were the peak of $\sigma$ is detected. The GRB afterglow location is consistent with the star formation peak region.

\section{Discussion}

\subsection{Jet break}

Although a full analysis of the temporal and spectral behaviour of the afterglow
is beyond the purpose of this paper, it is interesting to discuss some of the
main features. A break is observed simultaneously in the X-ray and optical light
curves. This has been traditionally interpreted in terms of a jetted outflow. In
this case, according to the standard afterglow model (\citealt{Zhang2004} and
references therein), the post-break decay slope is independent of the wavelength
and is determined from the electron energy distribution index.

For GRB\,091127, the X-ray light curve is better sampled and is not contaminated
by the SN and host components. However, the X-ray decay after the break is too
shallow even adopting an extreme value for the electron distribution index $p <
1$ (formally $p \approx 0.2$ is required, an unphysical and unpredented value),
which makes this model disfavored. The behaviour of the light curve after the
jet break is however determined by the complex physics that regulate the jet
sideways expansion, and is therefore quite uncertain
\citep{Granot2007,Meliani2010,van-Eerten2011}. Limiting our analysis before the jet break, our
data are roughly consistent with model predictions. A flat electron index is
required given the hard slope $\beta_{\rm X}$ of the X-ray spectrum ($p =
2\beta_{\rm X} = 1.86 \pm 0.20$). The optical and X-ray decay slopes predicted
for this value of $p$ are $\alpha_{\rm opt} = 0.72 \pm 0.04$, $\alpha_{\rm X} =
0.97 \pm 0.04$, compatible with the observed values within the errors (Table\,2). A low value of $p$ also predicts a blue optical spectrum with
$\beta_{\rm opt} = 0.43 \pm 0.10$, again consistent with the observed optical
SED.

A different possibility to explain the shallow decay of the optical and X-ray
light curves is long-lasting energy injection in the fireball (e.g.
\citealt{Zhang2006}), already invoked to explain several of the X-ray light curves
(e.g. \citealt{Panaitescu2006a,Liang2007}). In this case, the break might either pinpoint
the end of the energy injection, or represent the usual jet break, in which case
energy injection would last till after the end of the \textit{Swift}
observations. The former possibility is ruled out, since the observed $\alpha$
and $\beta$ in the X-ray range do not obey the expected closure relations
\citep{Zhang2004}. Even if energy injection is significant, we therefore conclude
that the most likely interpretation for the observed break is the jet effect.

\begin{figure}[!h]
   \centering
   \includegraphics[width=7.5cm, angle=90]{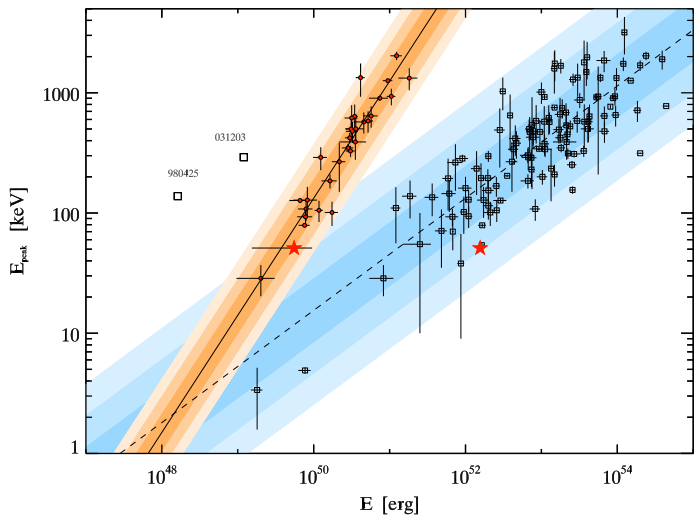}
      \caption{$E_{\rm peak}$--$E_{\rm iso}$ (empty squares) and $E_{\rm peak}$--$E_\gamma$ (red dots) correlations for GRBs.
      The shaded regions for the two correlations
represent the 1, 2, 3$\sigma$ scatter of the data points (computed perpendicularly) with respect to the best-fitted correlation (solid and dashed lines). The position of GRB\,091127 on both correlations is shown by the red stars. A color version of this figure is available online.}
         \label{ghirla}
   \end{figure}

Such interpretation is consistent with the typical break times of cosmological
GRBs (e.g. \citealt{Ghirlanda2005}). Using the {\it Fermi}-GBM spectral
parameters reported by \cite{Nava2011} (in agreement with those of the {\it
Fermi} team GCN; \citealt{Wilson-Hodge2009}) we show in Fig.\,\ref{ghirla} that
GRB\,091127 is consistent with the peak energy ($E_{\rm peak}$) - isotropic
energy ($E_{\rm iso}$) correlation and the $E_{\rm peak}$ - collimation correct energy
($E_{\gamma}$) correlation \citep{Amati2006,Ghirla2004}.

\subsection{Properties of the GRB\,091127 host}
The integrated properties of the host galaxy of GRB\,091127 (low mass, low luminosity, low metallicity and high SSFR) are comparable with those of the other host galaxies of LGRBs associated with SNe \citep{Sollerman2005,Starling2011} and more in general with LGRB hosts (e.g. \citealt{Savaglio2009,Levesque2010}). Our flux measurements implies an $A_V$ consistent with zero, in disagreement with the value obtained from the afterglow light-curve fitting. This indicates that the dust responsible for the GRB afterglow extinction is spatially localized and does not affect the overall emission of the galaxy.

 Recently, interest has grown on the mass-metallicity (\textit{M-Z}) relation of LGRB host galaxies. \cite{Levesque2010} (see also \citealt{Han2010}) showed that there is an offset between the {\it M-Z} relation of LGRB hosts with $z<0.3$ and that of nearby SDSS star-forming galaxies and between DEEP2 emission line galaxies at $\langle z \rangle=0.8$ and LGRB hosts with $0.3<z<1$. These works indicate that LGRBs occur preferentially in galaxies with lower metallicity compared to the general star-forming galaxy population. On the other hand, they also show that LGRB host galaxies do not follow a metallicity cut-off and therefore a low-metallicity host does not seem to be a key driver for the production of a LGRB. This result can be extended also to the LGRB environment, assuming that for such small galaxies there is a small metallicity gradient and that the average metallicity of the host is comparable to that of the GRB environment (e.g. \citealt{Levesque2011}). 

In a recent work, \cite{Mannucci2010} found a new general relation between the stellar mass, the metallicity and the SFR of galaxies (see also \citealt{Lara-Lopez2010}). This so-called fundamental metallicity relation (FMR) implies that, for a given $M_{\star}$, galaxies with higher SFR have lower metallicities and is explained by the authors by a complex interplay between infall gas dilution, SFR time-scale and gas outflows. \cite{Mannucci2011} use the FMR to investigate the {\it M-Z} relation of LGRB hosts and show that they follow the extrapolation of the FMR to low masses (see however \citealt{Kocevski2011}). This suggests that a key property of LGRB hosts could be the SSFR. Indeed there is a cut-off at log(SSFR)\,$\ge-10$ in the SSFR of LGRB hosts, indicating a high efficiency in forming stars.

    \begin{figure}[!h]
   \centering
   \includegraphics[width=9.0cm, angle=0]{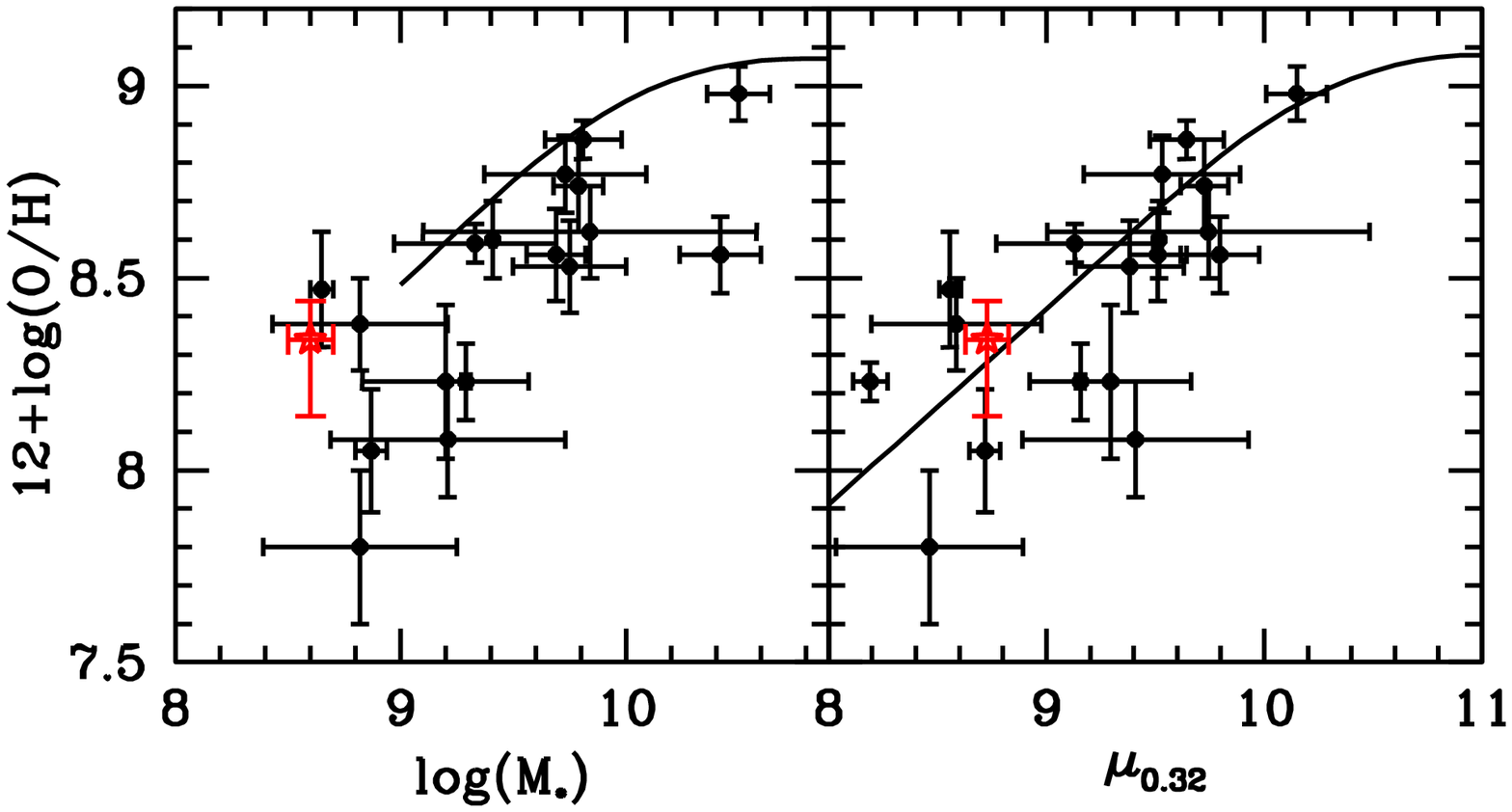}
      \caption{\textit{Left panel}: {\it M-Z} realtion. The black points with error bars correspond to the GRB $M_{\star}$ and metallicities used in \cite{Mannucci2011}, whereas the black line shows the fit of the local {\it M-Z} relation presented in the same paper. \textit{Right panel}: FMR relation. The black points with error bars correspond to the GRB $\mu_{0.32} \equiv \log(M_{\star})-0.32\log(\rm {SFR})$ and metallicities used in \cite{Mannucci2011}, whereas the black line shows the extended {\it M-Z} relation presented in the same paper. 
       In both panels the values for GRB\,091127 are represented by a red star with error bars. A color version of this figure is available online.
}
         \label{mannucci}
   \end{figure}

The host of GRB\,091127 extends to lower stellar masses the sample of LGRB host at $0.3<z<1$. The $M_{\star}$ and metallicity values we find broaden the space covered by the LGRB hosts in the {\it M-Z} plot presented by \cite{Levesque2010}, but can still be consistent with a {\it M-Z} shift of LGRB hosts compared with the {\it M-Z} relation found from emission lines galaxy surveys.
The SSFR value fulfills the above-mentioned cut-off and the $M_{\star}$, metallicity and SFR of the host of GRB\,091127 are in perfect agreement with the extended-FMR relation (see Fig.\,\ref{mannucci}). Moreover
we can use the FMR to predict the host metallicity starting from the $M_{\star}$ and SFR values determined. We use eq.\,2 by \cite{Mannucci2011} and we obtain 12+log(O/H)=8.3, in perfect agreement with our determined value. This exercise is an example of the potential of the FMR: if this relation is confirmed, it will be possible to determine all the three properties of galaxies even if the available observations are useful to determine two of them only.

We do not detect any Wolf-Rayet emission lines/bumps. Even though the detection of these features could be expected if Wolf-Rayet stars are the progenitors of LGRBs, the signal-to-noise and flux limits of our spectra make the detection not possible for lines with similar luminosities of those found by \cite{Han2010}.

The 2D velocity and velocity dispersion maps show that the kinematics of the GRB\,091127 host is complex. Following the works of \cite{Flores2006} and \cite{Puech2006} for a sample of galaxies at similar redshift studied with VLT/GIRAFFE, the perturbed kinematics could be associated with the presence of strong inflows/outflows or mergers (ongoing or remnant). Nonetheless it should be noted that the galaxies analysed by these authors are much more massive and luminous than LGRB hosts. 

The afterglow position is consistent with the region of the host characterised by high SFR, as expected from the findings of  \cite{Fruchter2006}. The peak of the SFR map is close to the position of the velocity dispersion maximum.
\cite{Green2010} using high resolution 3D observations, found that the high velocity dispersion is correlated with the star formation rates, claiming that star formation itself is the energetic driver of galaxy disk turbulence. Numerical simulations have already shown the same effect. In our case we are studying coarse maps and we only have information on a large scale. 
From our analysis it is not possible to assess a relation between the excess of SFR and the velocity dispersion behavior. Further modeling is necessary and a detailed analysis in this sense will be presented in a future work dedicated to the X-shooter GTO IFU LGRB host galaxy survey.

\section{Conclusions}
We have analysed the afterglow of GRB\,091127 using both \textit{Swift}-XRT and ground based data.
While the delay in the X-ray observations prevents us from studying the prompt/early emission characteristics as done in \cite{Starling2011}, we confirm the presence of a SN bump as reported by \cite{Cobb2010} (the GRB091127/SN\,2009nz association has been also recently claimed spectroscopically by \citealt{Berger2011}) and we point out the presence of a possible jet break, showing that GRB\,091127 follows the so-called Ghirlanda relation.

We carried out a detailed study of the properties of the host galaxy at $z\sim0.49$, using both slit and IFU X-shooter echelle spectroscopy. 2D GRB host maps have been previously produced only for the host of GRB\,980425 in the local universe \citep{Christensen2008}. 

Besides showing that the GRB\,091127 host has similar integrated properties as LGRB hosts in general, our results extend to lower values the $M_{\star}$ derived for the sample of LGRB hosts at $0.3<z<1.0$, therefore adding new information to explore the faint end of the {\it M-Z} relation.
 To have a conclusive picture on the shift of the LGRB host {\it M-Z} relation from that found from emission line galaxy surveys, it is necessary to increase the number of detailed studies of LGRB hosts.
The SFR, $M_{\star}$ and Z values determined are in perfect agreement with the FMR extension to low masses.

We stress the importance of LGRB hosts to study the properties of low-mass galaxies, since these galaxies are very often underrepresented in current surveys.
The {\it M-Z} is a fundamental relation to understand galaxy evolution. It is therefore very important to enlarge the sample of LGRB hosts to check if the {\it M-Z} offset is confirmed, to see how it evolves with redshift and to test the FMR, with the main purpose of understanding the processes that rule the {\it M-Z} relation.
Thanks to the X-shooter sensitivity and wavelength coverage, with the GTO program dedicated to GRB host spectroscopy we will be able to significantly enlarge the sample of studied LGRB host galaxies and to extend it to higher redshift.

IFU observations add detailed information to study this class 
of objects, not selected primarly by their luminosity and there-
fore forming a complementary sample to those of current surveys
of galaxies. 2D maps can also be used to understand which factors are required to drive a GRB explosion, and, for the cases with suitable spatial resolution, to determine the properties of the GRB region. 
The X-shooter IFU observation of the GRB 091127 host galaxy indicates a galaxy with 
perturbed rotation and a SFR enhancement towards the centre of
the host. The connection between turbulence and SFR is still debated.  
Thanks to the IFU X-shooter GTO LGRB host galaxy survey that we are carrying out within the French-Italian X-shooter GTO collaboration
it will
be possible to produce systematically 2D velocity, SFR and also
metallicity and electron density maps for more than 15 LGRB host galaxies at $z<0.6$. This study will provide a much detailed knowledge of the properties of these galaxies and of the GRB environment, to be compared with those of the star forming galaxy population and in particular with the host of broad-lined Ic SNe.

\begin{acknowledgements}
This work is partially based on observations with: Gran Telescopio Canarias (GTC), instaled in the Spanish Observatorio del Roque de los Muchachos of the Instituto de Astrof\'{\i}sica de Canarias in the island of La Palma; the Centro Astron\`omico Hispano Alem\'an (CAHA) at Calar Alto, operated jointly by the Max-Planck Institut f\"ur Astronomie and the Instituto de Astrof\'{\i}sica de Andaluc\'{\i}a (IAA-CSIC); the Nordic Optical Telescope, operated
on the island of La Palma jointly by Denmark, Finland, Iceland,
Norway, and Sweden, in the Spanish Observatorio del Roque de los
Muchachos of the Instituto de Astrof\'{\i}sica de Canarias; the APEX telescope under the ESO program E-084.D-0732A. 
This work made use of data supplied by the UK Swift Science Data Centre at the University of Leicester.
We thank the referee, Emily Levesque, for the accurate review of the paper.
SDV acknowledges useful correspondence with F. Mannucci, M. Modjaz and L. Nava. SDV thanks S. Savaglio and T. Vinci for precious help.
JG, RS, AJCT, JCT, MJ, acknowledge support by the Spanish Ministry of Science and Innovation (MICINN) under the project grants AYA2008-03467/ESP and AYA2009-14000-C03-01 (including Feder funds).
PJ acknowledges support by a Marie Curie European Reintegration Grant
within the 7th European Community Framework Program, and a Grant of
Excellence from the Icelandic Research Fund. The Dark Cosmology Centre is funded by the Danish National Research Foundation.
    \end{acknowledgements}

\bibliographystyle{aa} 
\bibliography{susy11} 

\begin{thebibliography}{90}
\expandafter\ifx\csname natexlab\endcsname\relax\def\natexlab#1{#1}\fi

\bibitem[{{Amati}(2006)}]{Amati2006}
{Amati}, L. 2006, \mnras, 372, 233

\bibitem[{{Baldry} \& {Glazebrook}(2003)}]{Baldry2003}
{Baldry}, I.~K. \& {Glazebrook}, K. 2003, \apj, 593, 258

\bibitem[{{Berger} {et~al.}(2011){Berger}, {Chornock}, {Holmes}, {Foley},
  {Cucchiara}, {Wolf}, {Podsiadlowski}, {Fox}, \& {Roth}}]{Berger2011}
{Berger}, E., {Chornock}, R., {Holmes}, T.~R., {et~al.} 2011, ApJ submitted,
  ArXiv:1106.3073

\bibitem[{{Bloom} {et~al.}(2002){Bloom}, {Kulkarni}, \&
  {Djorgovski}}]{Bloom2002}
{Bloom}, J.~S., {Kulkarni}, S.~R., \& {Djorgovski}, S.~G. 2002, \aj, 123, 1111

\bibitem[{{Bloom} {et~al.}(1999){Bloom}, {Kulkarni}, {Djorgovski},
  {Eichelberger}, {C{\^o}t{\'e}}, {Blakeslee}, {Odewahn}, {Harrison}, {Frail},
  {Filippenko}, {Leonard}, {Riess}, {Spinrad}, {Stern}, {Bunker}, {Dey},
  {Grossan}, {Perlmutter}, {Knop}, {Hook}, \& {Feroci}}]{Bloom1999}
{Bloom}, J.~S., {Kulkarni}, S.~R., {Djorgovski}, S.~G., {et~al.} 1999, \nat,
  401, 453

\bibitem[{{Bruzual}(2007)}]{Bruzual2007}
{Bruzual}, G. 2007, in Astronomical Society of the Pacific Conference Series,
  Vol. 374, From Stars to Galaxies: Building the Pieces to Build Up the
  Universe, ed. {A.~Vallenari, R.~Tantalo, L.~Portinari, \& A.~Moretti}, 303

\bibitem[{{Bruzual} \& {Charlot}(2003)}]{Bruzual2003}
{Bruzual}, G. \& {Charlot}, S. 2003, \mnras, 344, 1000

\bibitem[{{Calzetti} {et~al.}(1994){Calzetti}, {Kinney}, \&
  {Storchi-Bergmann}}]{Calzetti1994}
{Calzetti}, D., {Kinney}, A.~L., \& {Storchi-Bergmann}, T. 1994, \apj, 429, 582

\bibitem[{{Castro-Tirado} {et~al.}(2010){Castro-Tirado}, {M{\o}ller},
  {Garc{\'{\i}}a-Segura}, {Gorosabel}, {P{\'e}rez}, {de Ugarte Postigo},
  {Solano}, {Barrado}, {Klose}, {Kann}, {Castro Cer{\'o}n}, {Kouveliotou},
  {Fynbo}, {Hjorth}, {Pedersen}, {Pian}, {Rol}, {Palazzi}, {Masetti}, {Tanvir},
  {Vreeswijk}, {Andersen}, {Fruchter}, {Greiner}, {Wijers}, \& {van den
  Heuvel}}]{Castro-Tirado2010}
{Castro-Tirado}, A.~J., {M{\o}ller}, P., {Garc{\'{\i}}a-Segura}, G., {et~al.}
  2010, \aap, 517, A61

\bibitem[{{Chabrier}(2003)}]{Chabrier2003}
{Chabrier}, G. 2003, \pasp, 115, 763

\bibitem[{{Chen} {et~al.}(2007){Chen}, {Prochaska}, {Ramirez-Ruiz}, {Bloom},
  {Dessauges-Zavadsky}, \& {Foley}}]{Chen2007}
{Chen}, H.-W., {Prochaska}, J.~X., {Ramirez-Ruiz}, E., {et~al.} 2007, \apj,
  663, 420

\bibitem[{{Christensen} {et~al.}(2008){Christensen}, {Vreeswijk}, {Sollerman},
  {Th{\"o}ne}, {Le Floc'h}, \& {Wiersema}}]{Christensen2008}
{Christensen}, L., {Vreeswijk}, P.~M., {Sollerman}, J., {et~al.} 2008, \aap,
  490, 45

\bibitem[{{Cobb} {et~al.}(2010){Cobb}, {Bloom}, {Perley}, {Morgan}, {Cenko}, \&
  {Filippenko}}]{Cobb2010}
{Cobb}, B.~E., {Bloom}, J.~S., {Perley}, D.~A., {et~al.} 2010, \apjl, 718, L150

\bibitem[{{Covino} {et~al.}(2010){Covino}, {Campana}, {Conciatore}, {D'Elia},
  {Palazzi}, {Th{\"o}ne}, {Vergani}, {Wiersema}, {Brusasca}, {Cucchiara},
  {Cobb}, {Fern{\'a}ndez-Soto}, {Kann}, {Malesani}, {Tanvir}, {Antonelli},
  {Bremer}, {Castro-Tirado}, {de Ugarte Postigo}, {Molinari}, {Nicastro},
  {Stefanon}, {Testa}, {Tosti}, {Vitali}, {Amati}, {Chapman}, {Conconi},
  {Cutispoto}, {Fynbo}, {Goldoni}, {Henriksen}, {Horne}, {Malaspina}, {Meurs},
  {Pian}, {Stella}, {Tagliaferri}, {Ward}, \& {Zerbi}}]{Covino2010}
{Covino}, S., {Campana}, S., {Conciatore}, M.~L., {et~al.} 2010, \aap, 521, A53

\bibitem[{{Crowther}(2007)}]{Crowther2007}
{Crowther}, P.~A. 2007, \araa, 45, 177

\bibitem[{{Cucchiara} {et~al.}(2009){Cucchiara}, {Fox}, {Levan}, \&
  {Tanvir}}]{Cucchiara2009a}
{Cucchiara}, A., {Fox}, D., {Levan}, A., \& {Tanvir}, N. 2009, GCN10202

\bibitem[{{Dahlen} {et~al.}(2005){Dahlen}, {Mobasher}, {Somerville},
  {Moustakas}, {Dickinson}, {Ferguson}, \& {Giavalisco}}]{Dahlen2005}
{Dahlen}, T., {Mobasher}, B., {Somerville}, R.~S., {et~al.} 2005, \apj, 631,
  126

\bibitem[{{Della Valle} {et~al.}(2006){Della Valle}, {Chincarini}, {Panagia},
  {Tagliaferri}, {Malesani}, {Testa}, {Fugazza}, {Campana}, {Covino},
  {Mangano}, {Antonelli}, {D'Avanzo}, {Hurley}, {Mirabel}, {Pellizza},
  {Piranomonte}, \& {Stella}}]{Della-Valle2006}
{Della Valle}, M., {Chincarini}, G., {Panagia}, N., {et~al.} 2006, \nat, 444,
  1050

\bibitem[{{Della Valle} {et~al.}(2003){Della Valle}, {Malesani}, {Benetti},
  {Testa}, {Hamuy}, {Antonelli}, {Chincarini}, {Cocozza}, {Covino}, {D'Avanzo},
  {Fugazza}, {Ghisellini}, {Gilmozzi}, {Lazzati}, {Mason}, {Mazzali}, \&
  {Stella}}]{Della-Valle2003}
{Della Valle}, M., {Malesani}, D., {Benetti}, S., {et~al.} 2003, \aap, 406, L33

\bibitem[{{di Paola}(2001)}]{di-Paola2001}
{di Paola}, A. 2001, in Gamma-ray Bursts in the Afterglow Era, ed. {E.~Costa,
  F.~Frontera, \& J.~Hjorth}, 390

\bibitem[{{D'Odorico} {et~al.}(2004){D'Odorico}, {Andersen}, {Conconi}, {De
  Caprio}, {Delabre}, {Di Marcantonio}, {Dekker}, {Downing}, {Finger}, {Groot},
  {Hanenburg}, {Hammer}, {Horville}, {Hjorth}, {Kaper}, {Klougart},
  {Kjaergaard-Rasmussen}, {Lizon}, {Marteaud}, {Mazzoleni}, {Michaelsen},
  {Pallavicini}, {Rigal}, {Santin}, {Norup Soerensen}, {Spano}, {Venema},
  {Vola}, \& {Zerbi}}]{DOdorico2004}
{D'Odorico}, S., {Andersen}, M.~I., {Conconi}, P., {et~al.} 2004, in Society of
  Photo-Optical Instrumentation Engineers (SPIE) Conference Series, Vol. 5492,
  Society of Photo-Optical Instrumentation Engineers (SPIE) Conference Series,
  ed. {A.~F.~M.~Moorwood \& M.~Iye}, 220--229

\bibitem[{{Evans} {et~al.}(2009){Evans}, {Beardmore}, {Page}, {Osborne},
  {O'Brien}, {Willingale}, {Starling}, {Burrows}, {Godet}, {Vetere}, {Racusin},
  {Goad}, {Wiersema}, {Angelini}, {Capalbi}, {Chincarini}, {Gehrels}, {Kennea},
  {Margutti}, {Morris}, {Mountford}, {Pagani}, {Perri}, {Romano}, \&
  {Tanvir}}]{Evans2009}
{Evans}, P.~A., {Beardmore}, A.~P., {Page}, K.~L., {et~al.} 2009, \mnras, 397,
  1177

\bibitem[{{Evans} {et~al.}(2007){Evans}, {Beardmore}, {Page}, {Tyler},
  {Osborne}, {Goad}, {O'Brien}, {Vetere}, {Racusin}, {Morris}, {Burrows},
  {Capalbi}, {Perri}, {Gehrels}, \& {Romano}}]{Evans2007}
{Evans}, P.~A., {Beardmore}, A.~P., {Page}, K.~L., {et~al.} 2007, \aap, 469,
  379

\bibitem[{{Fiore} {et~al.}(2005){Fiore}, {D'Elia}, {Lazzati}, {Perna},
  {Sbordone}, {Stratta}, {Meurs}, {Ward}, {Antonelli}, {Chincarini}, {Covino},
  {Di Paola}, {Fontana}, {Ghisellini}, {Israel}, {Frontera}, {Marconi},
  {Stella}, {Vietri}, \& {Zerbi}}]{Fiore2005}
{Fiore}, F., {D'Elia}, V., {Lazzati}, D., {et~al.} 2005, \apj, 624, 853

\bibitem[{{Flores} {et~al.}(2011){Flores}, {Goldoni}, {Royer}, {Piranomonte},
  {Vergani}, {Onori}, {Palazzi}, {Covino}, {Randich}, {Hammer}, {Pian},
  {Savaglio}, \& {Tagliaferri}}]{Flores2011}
{Flores}, H., {Goldoni}, P., {Royer}, F., {et~al.} 2011, Astronomische
  Nachrichten, 332, 288

\bibitem[{{Flores} {et~al.}(2006){Flores}, {Hammer}, {Puech}, {Amram}, \&
  {Balkowski}}]{Flores2006}
{Flores}, H., {Hammer}, F., {Puech}, M., {Amram}, P., \& {Balkowski}, C. 2006,
  \aap, 455, 107

\bibitem[{{Fruchter} {et~al.}(2006){Fruchter}, {Levan}, {Strolger},
  {Vreeswijk}, {Thorsett}, {Bersier}, {Burud}, {Castro Cer{\'o}n},
  {Castro-Tirado}, {Conselice}, {Dahlen}, {Ferguson}, {Fynbo}, {Garnavich},
  {Gibbons}, {Gorosabel}, {Gull}, {Hjorth}, {Holland}, {Kouveliotou}, {Levay},
  {Livio}, {Metzger}, {Nugent}, {Petro}, {Pian}, {Rhoads}, {Riess}, {Sahu},
  {Smette}, {Tanvir}, {Wijers}, \& {Woosley}}]{Fruchter2006}
{Fruchter}, A.~S., {Levan}, A.~J., {Strolger}, L., {et~al.} 2006, \nat, 441,
  463

\bibitem[{{Fynbo} {et~al.}(2006){Fynbo}, {Watson}, {Th{\"o}ne}, {Sollerman},
  {Bloom}, {Davis}, {Hjorth}, {Jakobsson}, {J{\o}rgensen}, {Graham},
  {Fruchter}, {Bersier}, {Kewley}, {Cassan}, {Castro Cer{\'o}n}, {Foley},
  {Gorosabel}, {Hinse}, {Horne}, {Jensen}, {Klose}, {Kocevski}, {Marquette},
  {Perley}, {Ramirez-Ruiz}, {Stritzinger}, {Vreeswijk}, {Wijers}, {Woller},
  {Xu}, \& {Zub}}]{Fynbo2006}
{Fynbo}, J.~P.~U., {Watson}, D., {Th{\"o}ne}, C.~C., {et~al.} 2006, \nat, 444,
  1047

\bibitem[{{Gal-Yam} {et~al.}(2006){Gal-Yam}, {Fox}, {Price}, {Ofek}, {Davis},
  {Leonard}, {Soderberg}, {Schmidt}, {Lewis}, {Peterson}, {Kulkarni}, {Berger},
  {Cenko}, {Sari}, {Sharon}, {Frail}, {Moon}, {Brown}, {Cucchiara}, {Harrison},
  {Piran}, {Persson}, {McCarthy}, {Penprase}, {Chevalier}, \&
  {MacFadyen}}]{Gal-Yam2006}
{Gal-Yam}, A., {Fox}, D.~B., {Price}, P.~A., {et~al.} 2006, \nat, 444, 1053

\bibitem[{{Galama} {et~al.}(1998){Galama}, {Groot}, {van Paradijs},
  {Kouveliotou}, {Strom}, {Wijers}, {Tanvir}, {Bloom}, {Centurion}, {Telting},
  {Rutten}, {Smith}, {Mackey}, {Smartt}, {Benn}, {Heise}, \& {in 't
  Zand}}]{Galama1998}
{Galama}, T.~J., {Groot}, P.~J., {van Paradijs}, J., {et~al.} 1998, \apjl, 497,
  L13

\bibitem[{{Galama} {et~al.}(2000){Galama}, {Tanvir}, {Vreeswijk}, {Wijers},
  {Groot}, {Rol}, {van Paradijs}, {Kouveliotou}, {Fruchter}, {Masetti},
  {Pedersen}, {Margon}, {Deutsch}, {Metzger}, {Armus}, {Klose}, \&
  {Stecklum}}]{Galama2000}
{Galama}, T.~J., {Tanvir}, N., {Vreeswijk}, P.~M., {et~al.} 2000, \apj, 536,
  185

\bibitem[{{Ghirlanda} {et~al.}(2005){Ghirlanda}, {Ghisellini}, \&
  {Firmani}}]{Ghirlanda2005}
{Ghirlanda}, G., {Ghisellini}, G., \& {Firmani}, C. 2005, \mnras, 361, L10

\bibitem[{{Ghirlanda} {et~al.}(2004){Ghirlanda}, {Ghisellini}, \&
  {Lazzati}}]{Ghirla2004}
{Ghirlanda}, G., {Ghisellini}, G., \& {Lazzati}, D. 2004, \apj, 616, 331

\bibitem[{{Goldoni} {et~al.}(2006){Goldoni}, {Royer}, {Fran{\c c}ois},
  {Horrobin}, {Blanc}, {Vernet}, {Modigliani}, \& {Larsen}}]{Goldoni2006}
{Goldoni}, P., {Royer}, F., {Fran{\c c}ois}, P., {et~al.} 2006, in Society of
  Photo-Optical Instrumentation Engineers (SPIE) Conference Series, Vol. 6269,
  Society of Photo-Optical Instrumentation Engineers (SPIE) Conference Series

\bibitem[{{Granot}(2007)}]{Granot2007}
{Granot}, J. 2007, in Revista Mexicana de Astronomia y Astrofisica Conference
  Series, Vol.~27, Revista Mexicana de Astronomia y Astrofisica, vol. 27,
  140--165

\bibitem[{{Green} {et~al.}(2010){Green}, {Glazebrook}, {McGregor}, {Abraham},
  {Poole}, {Damjanov}, {McCarthy}, {Colless}, \& {Sharp}}]{Green2010}
{Green}, A.~W., {Glazebrook}, K., {McGregor}, P.~J., {et~al.} 2010, \nat, 467,
  684

\bibitem[{{Guetta} \& {Della Valle}(2007)}]{Guetta2007}
{Guetta}, D. \& {Della Valle}, M. 2007, \apjl, 657, L73

\bibitem[{{Hammer} {et~al.}(2006){Hammer}, {Flores}, {Schaerer},
  {Dessauges-Zavadsky}, {Le Floc'h}, \& {Puech}}]{Hammer2006}
{Hammer}, F., {Flores}, H., {Schaerer}, D., {et~al.} 2006, \aap, 454, 103

\bibitem[{{Han} {et~al.}(2010){Han}, {Hammer}, {Liang}, {Flores}, {Rodrigues},
  {Hou}, \& {Wei}}]{Han2010}
{Han}, X.~H., {Hammer}, F., {Liang}, Y.~C., {et~al.} 2010, \aap, 514, A24

\bibitem[{{Heger} {et~al.}(2003){Heger}, {Fryer}, {Woosley}, {Langer}, \&
  {Hartmann}}]{Heger2003}
{Heger}, A., {Fryer}, C.~L., {Woosley}, S.~E., {Langer}, N., \& {Hartmann},
  D.~H. 2003, \apj, 591, 288

\bibitem[{{Hjorth} \& {Bloom}(2011)}]{Hjorth2011}
{Hjorth}, J. \& {Bloom}, J.~S. 2011, Chapter 9 in "Gamma-Ray Bursts", eds. C.
  Kouveliotou, R. A. M. J. Wijers, S. E. Woosley, Cambridge University Press,
  ArXiv:1104.2274

\bibitem[{{Kelly} {et~al.}(2008){Kelly}, {Kirshner}, \& {Pahre}}]{Kelly2008}
{Kelly}, P.~L., {Kirshner}, R.~P., \& {Pahre}, M. 2008, \apj, 687, 1201

\bibitem[{{Kennicutt}(1998)}]{Kennicutt1998}
{Kennicutt}, Jr., R.~C. 1998, \araa, 36, 189

\bibitem[{{Kewley} \& {Ellison}(2008)}]{Kewley2008}
{Kewley}, L.~J. \& {Ellison}, S.~L. 2008, \apj, 681, 1183

\bibitem[{{Kobulnicky} \& {Kewley}(2004)}]{Kobulnicky2004}
{Kobulnicky}, H.~A. \& {Kewley}, L.~J. 2004, \apj, 617, 240

\bibitem[{{Kocevski} \& {West}(2011)}]{Kocevski2011}
{Kocevski}, D. \& {West}, A.~A. 2011, \apjl, 735, L8+

\bibitem[{{Lara-L{\'o}pez} {et~al.}(2010){Lara-L{\'o}pez}, {Cepa},
  {Bongiovanni}, {P{\'e}rez Garc{\'{\i}}a}, {Ederoclite}, {Casta{\~n}eda},
  {Fern{\'a}ndez Lorenzo}, {Povi{\'c}}, \&
  {S{\'a}nchez-Portal}}]{Lara-Lopez2010}
{Lara-L{\'o}pez}, M.~A., {Cepa}, J., {Bongiovanni}, A., {et~al.} 2010, \aap,
  521, L53

\bibitem[{{Lazzati} {et~al.}(2001){Lazzati}, {Covino}, {Ghisellini}, {Fugazza},
  {Campana}, {Saracco}, {Price}, {Berger}, {Kulkarni}, {Ramirez-Ruiz},
  {Cimatti}, {Della Valle}, {di Serego Alighieri}, {Celotti}, {Haardt},
  {Israel}, \& {Stella}}]{Lazzati2001a}
{Lazzati}, D., {Covino}, S., {Ghisellini}, G., {et~al.} 2001, \aap, 378, 996

\bibitem[{{Levesque} {et~al.}(2011){Levesque}, {Berger}, {Soderberg}, \&
  {Chornock}}]{Levesque2011}
{Levesque}, E.~M., {Berger}, E., {Soderberg}, A.~M., \& {Chornock}, R. 2011,
  ApJ submitted, ArXiv:1104.2865

\bibitem[{{Levesque} {et~al.}(2010){Levesque}, {Kewley}, {Berger}, \& {Jabran
  Zahid}}]{Levesque2010}
{Levesque}, E.~M., {Kewley}, L.~J., {Berger}, E., \& {Jabran Zahid}, H. 2010,
  \aj, 140, 1557

\bibitem[{{Liang} {et~al.}(2007){Liang}, {Zhang}, \& {Zhang}}]{Liang2007}
{Liang}, E.-W., {Zhang}, B.-B., \& {Zhang}, B. 2007, \apj, 670, 565

\bibitem[{{MacFadyen} \& {Woosley}(1999)}]{MacFadyen1999}
{MacFadyen}, A.~I. \& {Woosley}, S.~E. 1999, \apj, 524, 262

\bibitem[{{Maiolino} {et~al.}(2008){Maiolino}, {Nagao}, {Grazian}, {Cocchia},
  {Marconi}, {Mannucci}, {Cimatti}, {Pipino}, {Ballero}, {Calura}, {Chiappini},
  {Fontana}, {Granato}, {Matteucci}, {Pastorini}, {Pentericci}, {Risaliti},
  {Salvati}, \& {Silva}}]{Maiolino2008}
{Maiolino}, R., {Nagao}, T., {Grazian}, A., {et~al.} 2008, \aap, 488, 463

\bibitem[{{Mannucci} {et~al.}(2010){Mannucci}, {Cresci}, {Maiolino}, {Marconi},
  \& {Gnerucci}}]{Mannucci2010}
{Mannucci}, F., {Cresci}, G., {Maiolino}, R., {Marconi}, A., \& {Gnerucci}, A.
  2010, \mnras, 408, 2115

\bibitem[{{Mannucci} {et~al.}(2011){Mannucci}, {Salvaterra}, \&
  {Campisi}}]{Mannucci2011}
{Mannucci}, F., {Salvaterra}, R., \& {Campisi}, M.~A. 2011, \mnras, 439

\bibitem[{{McKenzie} \& {Schaefer}(1999)}]{McKenzie1999}
{McKenzie}, E.~H. \& {Schaefer}, B.~E. 1999, \pasp, 111, 964

\bibitem[{{Meliani} \& {Keppens}(2010)}]{Meliani2010}
{Meliani}, Z. \& {Keppens}, R. 2010, \aap, 520, L3

\bibitem[{{Mirabal} {et~al.}(2003){Mirabal}, {Halpern}, {Chornock},
  {Filippenko}, {Terndrup}, {Armstrong}, {Kemp}, {Thorstensen}, {Tavarez}, \&
  {Espaillat}}]{Mirabal2003}
{Mirabal}, N., {Halpern}, J.~P., {Chornock}, R., {et~al.} 2003, \apj, 595, 935

\bibitem[{{Modjaz} {et~al.}(2008){Modjaz}, {Kewley}, {Kirshner}, {Stanek},
  {Challis}, {Garnavich}, {Greene}, {Kelly}, \& {Prieto}}]{Modjaz2008}
{Modjaz}, M., {Kewley}, L., {Kirshner}, R.~P., {et~al.} 2008, \aj, 135, 1136

\bibitem[{{Morrison} \& {McCammon}(1983)}]{Morrison1983}
{Morrison}, R. \& {McCammon}, D. 1983, \apj, 270, 119

\bibitem[{{Nava} {et~al.}(2011){Nava}, {Ghirlanda}, {Ghisellini}, \&
  {Celotti}}]{Nava2011}
{Nava}, L., {Ghirlanda}, G., {Ghisellini}, G., \& {Celotti}, A. 2011, \aap,
  530, A21

\bibitem[{{Osterbrock}(1989)}]{Osterbrock1989}
{Osterbrock}, D.~E. 1989, {Astrophysics of gaseous nebulae and active galactic
  nuclei} (Mill Valley, CA, University Science Books)

\bibitem[{{Panaitescu} {et~al.}(2006){Panaitescu}, {M{\'e}sz{\'a}ros},
  {Gehrels}, {Burrows}, \& {Nousek}}]{Panaitescu2006a}
{Panaitescu}, A., {M{\'e}sz{\'a}ros}, P., {Gehrels}, N., {Burrows}, D., \&
  {Nousek}, J. 2006, \mnras, 366, 1357

\bibitem[{{Patat} {et~al.}(2001){Patat}, {Cappellaro}, {Danziger}, {Mazzali},
  {Sollerman}, {Augusteijn}, {Brewer}, {Doublier}, {Gonzalez}, {Hainaut},
  {Lidman}, {Leibundgut}, {Nomoto}, {Nakamura}, {Spyromilio}, {Rizzi},
  {Turatto}, {Walsh}, {Galama}, {van Paradijs}, {Kouveliotou}, {Vreeswijk},
  {Frontera}, {Masetti}, {Palazzi}, \& {Pian}}]{Patat2001}
{Patat}, F., {Cappellaro}, E., {Danziger}, J., {et~al.} 2001, \apj, 555, 900

\bibitem[{{Pei}(1992)}]{Pei1992}
{Pei}, Y.~C. 1992, \apj, 395, 130

\bibitem[{{Pettini} \& {Pagel}(2004)}]{Pettini2004a}
{Pettini}, M. \& {Pagel}, B.~E.~J. 2004, \mnras, 348, L59

\bibitem[{{Puech} {et~al.}(2006){Puech}, {Hammer}, {Flores}, {{\"O}stlin}, \&
  {Marquart}}]{Puech2006}
{Puech}, M., {Hammer}, F., {Flores}, H., {{\"O}stlin}, G., \& {Marquart}, T.
  2006, \aap, 455, 119

\bibitem[{{Salpeter}(1955)}]{Salpeter1955}
{Salpeter}, E.~E. 1955, \apj, 121, 161

\bibitem[{{Savaglio} {et~al.}(2009){Savaglio}, {Glazebrook}, \& {Le
  Borgne}}]{Savaglio2009}
{Savaglio}, S., {Glazebrook}, K., \& {Le Borgne}, D. 2009, \apj, 691, 182

\bibitem[{{Schaefer} {et~al.}(2003){Schaefer}, {Gerardy}, {H{\"o}flich},
  {Panaitescu}, {Quimby}, {Mader}, {Hill}, {Kumar}, {Wheeler}, {Eracleous},
  {Sigurdsson}, {M{\'e}sz{\'a}ros}, {Zhang}, {Wang}, {Hessman}, \&
  {Petrosian}}]{Schaefer2003}
{Schaefer}, B.~E., {Gerardy}, C.~L., {H{\"o}flich}, P., {et~al.} 2003, \apj,
  588, 387

\bibitem[{{Schlegel} {et~al.}(1998){Schlegel}, {Finkbeiner}, \&
  {Davis}}]{Schlegel1998}
{Schlegel}, D.~J., {Finkbeiner}, D.~P., \& {Davis}, M. 1998, \apj, 500, 525

\bibitem[{{Skrutskie} {et~al.}(2006){Skrutskie}, {Cutri}, {Stiening},
  {Weinberg}, {Schneider}, {Carpenter}, {Beichman}, {Capps}, {Chester},
  {Elias}, {Huchra}, {Liebert}, {Lonsdale}, {Monet}, {Price}, {Seitzer},
  {Jarrett}, {Kirkpatrick}, {Gizis}, {Howard}, {Evans}, {Fowler}, {Fullmer},
  {Hurt}, {Light}, {Kopan}, {Marsh}, {McCallon}, {Tam}, {Van Dyk}, \&
  {Wheelock}}]{Skrutskie2006}
{Skrutskie}, M.~F., {Cutri}, R.~M., {Stiening}, R., {et~al.} 2006, \aj, 131,
  1163

\bibitem[{{Soderberg} {et~al.}(2010){Soderberg}, {Chakraborti}, {Pignata},
  {Chevalier}, {Chandra}, {Ray}, {Wieringa}, {Copete}, {Chaplin},
  {Connaughton}, {Barthelmy}, {Bietenholz}, {Chugai}, {Stritzinger}, {Hamuy},
  {Fransson}, {Fox}, {Levesque}, {Grindlay}, {Challis}, {Foley}, {Kirshner},
  {Milne}, \& {Torres}}]{Soderberg2010}
{Soderberg}, A.~M., {Chakraborti}, S., {Pignata}, G., {et~al.} 2010, \nat, 463,
  513

\bibitem[{{Sollerman} {et~al.}(2002){Sollerman}, {Holland}, {Challis},
  {Fransson}, {Garnavich}, {Kirshner}, {Kozma}, {Leibundgut}, {Lundqvist},
  {Patat}, {Filippenko}, {Panagia}, \& {Wheeler}}]{Sollerman2002}
{Sollerman}, J., {Holland}, S.~T., {Challis}, P., {et~al.} 2002, \aap, 386, 944

\bibitem[{{Sollerman} {et~al.}(2000){Sollerman}, {Kozma}, {Fransson},
  {Leibundgut}, {Lundqvist}, {Ryde}, \& {Woudt}}]{Sollerman2000}
{Sollerman}, J., {Kozma}, C., {Fransson}, C., {et~al.} 2000, \apjl, 537, L127

\bibitem[{{Sollerman} {et~al.}(2005){Sollerman}, {{\"O}stlin}, {Fynbo},
  {Hjorth}, {Fruchter}, \& {Pedersen}}]{Sollerman2005}
{Sollerman}, J., {{\"O}stlin}, G., {Fynbo}, J.~P.~U., {et~al.} 2005, New
  Astronomy, 11, 103

\bibitem[{{Sparre} {et~al.}(2011){Sparre}, {Sollerman}, {Fynbo}, {Malesani},
  {Goldoni}, {de Ugarte Postigo}, {Covino}, {D'Elia}, {Flores}, {Hammer},
  {Hjorth}, {Jakobsson}, {Kaper}, {Leloudas}, {Levan}, {Milvang-Jensen},
  {Schulze}, {Tagliaferri}, {Tanvir}, {Watson}, {Wiersema}, \&
  {Wijers}}]{Sparre2011}
{Sparre}, M., {Sollerman}, J., {Fynbo}, J.~P.~U., {et~al.} 2011, \apjl, 735,
  L24

\bibitem[{{Starling} {et~al.}(2011){Starling}, {Wiersema}, {Levan}, {Sakamoto},
  {Bersier}, {Goldoni}, {Oates}, {Rowlinson}, {Campana}, {Sollerman}, {Tanvir},
  {Malesani}, {Fynbo}, {Covino}, {D'Avanzo}, {O'Brien}, {Page}, {Osborne},
  {Vergani}, {Barthelmy}, {Burrows}, {Cano}, {Curran}, {de Pasquale}, {D'Elia},
  {Evans}, {Flores}, {Fruchter}, {Garnavich}, {Gehrels}, {Gorosabel}, {Hjorth},
  {Holland}, {van der Horst}, {Hurkett}, {Jakobsson}, {Kamble}, {Kouveliotou},
  {Kuin}, {Kaper}, {Mazzali}, {Nugent}, {Pian}, {Stamatikos}, {Th{\"o}ne}, \&
  {Woosley}}]{Starling2011}
{Starling}, R.~L.~C., {Wiersema}, K., {Levan}, A.~J., {et~al.} 2011, \mnras,
  411, 2792

\bibitem[{{Starling} {et~al.}(2005){Starling}, {Wijers}, {Hughes}, {Tanvir},
  {Vreeswijk}, {Rol}, \& {Salamanca}}]{Starling2005a}
{Starling}, R.~L.~C., {Wijers}, R.~A.~M.~J., {Hughes}, M.~A., {et~al.} 2005,
  \mnras, 360, 305

\bibitem[{{Th{\"o}ne} {et~al.}(2008){Th{\"o}ne}, {Fynbo}, {{\"O}stlin},
  {Milvang-Jensen}, {Wiersema}, {Malesani}, {Della Monica Ferreira},
  {Gorosabel}, {Kann}, {Watson}, {Micha{\l}owski}, {Fruchter}, {Levan},
  {Hjorth}, \& {Sollerman}}]{Thone2008a}
{Th{\"o}ne}, C.~C., {Fynbo}, J.~P.~U., {{\"O}stlin}, G., {et~al.} 2008, \apj,
  676, 1151

\bibitem[{{Th{\"o}ne} {et~al.}(2009){Th{\"o}ne}, {Goldoni}, {Covino},
  {Antonelli}, {Malesani}, {Fynbo}, {Levan}, {Jakobsson}, {Flores},
  {Milvang-Jensen}, {Hjorth}, {Watson}, {Wiersema}, {Tanvir}, \& {de Ugarte
  Postigo}}]{Thone2009}
{Th{\"o}ne}, C.~C., {Goldoni}, P., {Covino}, S., {et~al.} 2009, GCN 10233, 233,
  1

\bibitem[{{van Dokkum}(2001)}]{van-Dokkum2001}
{van Dokkum}, P.~G. 2001, \pasp, 113, 1420

\bibitem[{{van Eerten} \& {MacFadyen}(2011)}]{van-Eerten2011}
{van Eerten}, H.~J. \& {MacFadyen}, A.~I. 2011, ArXiv e-prints

\bibitem[{{van Marle} {et~al.}(2008){van Marle}, {Langer}, {Yoon}, \&
  {Garc{\'{\i}}a-Segura}}]{van-Marle2008}
{van Marle}, A.~J., {Langer}, N., {Yoon}, S.-C., \& {Garc{\'{\i}}a-Segura}, G.
  2008, \aap, 478, 769

\bibitem[{{Wilson-Hodge} \& {Preece}(2009)}]{Wilson-Hodge2009}
{Wilson-Hodge}, C.~A. \& {Preece}, R.~D. 2009, GCN 10204, 204, 1

\bibitem[{{Woosley}(1993)}]{Woosley1993}
{Woosley}, S.~E. 1993, \apj, 405, 273

\bibitem[{{Yang} {et~al.}(2008){Yang}, {Flores}, {Hammer}, {Neichel}, {Puech},
  {Nesvadba}, {Rawat}, {Cesarsky}, {Lehnert}, {Pozzetti}, {Fuentes-Carrera},
  {Amram}, {Balkowski}, {Dannerbauer}, {di Serego Alighieri}, {Guiderdoni},
  {Kembhavi}, {Liang}, {{\"O}stlin}, {Ravikumar}, {Vergani}, {Vernet}, \&
  {Wozniak}}]{Yang2008}
{Yang}, Y., {Flores}, H., {Hammer}, F., {et~al.} 2008, \aap, 477, 789

\bibitem[{{Yoon} {et~al.}(2010){Yoon}, {Woosley}, \& {Langer}}]{Yoon2010}
{Yoon}, S.-C., {Woosley}, S.~E., \& {Langer}, N. 2010, \apj, 725, 940

\bibitem[{{Zhang} {et~al.}(2006){Zhang}, {Fan}, {Dyks}, {Kobayashi},
  {M{\'e}sz{\'a}ros}, {Burrows}, {Nousek}, \& {Gehrels}}]{Zhang2006}
{Zhang}, B., {Fan}, Y.~Z., {Dyks}, J., {et~al.} 2006, \apj, 642, 354

\bibitem[{{Zhang} \& {M{\'e}sz{\'a}ros}(2004)}]{Zhang2004}
{Zhang}, B. \& {M{\'e}sz{\'a}ros}, P. 2004, International Journal of Modern
  Physics A, 19, 2385

\end{thebibliography}

\end{document}